\pdfoutput=1

\documentclass[11pt,twoside,a4paper,cmspaper,final,collab]{cms-tdr}
\def\svnVersion{f733fbe-D}\def\svnDate{2020/02/26}\def\cmsCernNoTag{CERN-EP-2019-189}\def\cmsCernDate{\today}\def\cmsMessage{"Published in Physics Letters B as \href{http://dx.doi.org/10.1016/j.physletb.2020.135263}{\doi{10.1016/j.physletb.2020.135263}.}"}

\begin{document}\cmsNoteHeader{TOP-19-007}

\hyphenation{had-ron-i-za-tion}
\hyphenation{cal-or-i-me-ter}
\hyphenation{de-vices}

\newcommand{\pp}{\Pp{}\Pp{}\xspace}
\newcommand{\emu}{\Pe{}$^{\pm}$\Pgm{}$^{\mp}$\xspace}
\newcommand{\Minuit} {{\textsc{Minuit}}\xspace}
\newcommand{\Minos} {{\textsc{Minos}}\xspace}
\newcommand{\Hathor} {{\textsc{Hathor}}\xspace}
\newcommand{\RunDec} {{\textsc{CRunDec}}\xspace}
\newcommand{\Wjets}{\PW{}+jets\xspace}
\newcommand{\intlumi}{\ensuremath{\mathcal{L}}\xspace}
\newcommand{\agreement}{1.1\xspace}
\newcommand{\ratiocorronetwo}{13\%\xspace}
\newcommand{\ratiocorronethree}{\ensuremath{-}45\%\xspace}
\newcommand{\ratiocorrtwothree}{11\%\xspace}

\newcommand{\rmu}{\ensuremath{r(\mu)}\xspace}
\newcommand{\rhoik}{\ensuremath{\rho_{ik}}\xspace}
\newcommand{\muk}{\ensuremath{\mu_k}\xspace}
\newcommand{\xmin}{\ensuremath{\hat{x}}\xspace}
\newcommand{\muref}{\ensuremath{\mu_\text{ref}}\xspace}
\newcommand{\nf}{\ensuremath{n_f}\xspace}
\newcommand{\as}{\ensuremath{\alpS}\xspace}
\newcommand{\asmz}{\ensuremath{\as(m_{\cPZ})}\xspace}
\newcommand{\mur}{\ensuremath{\mu_\mathrm{r}}\xspace}
\newcommand{\muf}{\ensuremath{\mu_\mathrm{f}}\xspace}
\newcommand{\stt}{\ensuremath{\sigma_\ttbar}\xspace}
\newcommand{\dstt}{\ensuremath{\rd\stt/\rd\mtt}\xspace}
\newcommand{\sttk}{\ensuremath{\sigma_\ttbar^{(\mu_k)}}\xspace}
\newcommand{\msbar}{\ensuremath{\mathrm{\overline{MS}}}\xspace}
\newcommand{\mt}{\ensuremath{m_{\cPqt}}\xspace}
\newcommand{\mtt}{\ensuremath{m_\ttbar}\xspace}
\newcommand{\Dmtt}{\ensuremath{\Delta \mtt^k}\xspace}
\newcommand{\Nev}{\ensuremath{N_{\text{events}}}\xspace}
\newcommand{\Njets}{\ensuremath{N_{\text{jets}}}\xspace}
\newcommand{\Nb}{\ensuremath{N_{\cPqb}}\xspace}
\newcommand{\tW}{\ensuremath{\cPqt\PW}\xspace}

\newcommand{\mttreco}{\ensuremath{m_\ttbar^{\text{reco}}}\xspace}
\newcommand{\mtmt}{\ensuremath{\mt(\mt)}\xspace}
\newcommand{\mtot}{\ensuremath{\mt^\text{incl}(\mt)}\xspace}
\newcommand{\mtmu}{\ensuremath{\mt(\mu)}\xspace}
\newcommand{\mtmuk}{\ensuremath{\mt(\muk)}\xspace}
\newcommand{\mtmutwo}{\ensuremath{\mt(\mu_2)}\xspace}
\newcommand{\mtmuref}{\ensuremath{\mt(\muref)}\xspace}

\newcommand{\mtmc}{\ensuremath{\mt^{\mathrm{MC}}}\xspace}
\newcommand{\mtkin}{\ensuremath{\mt^{\text{kin}}}\xspace}
\newcommand{\mlb}{\ensuremath{m_{\ell\cPqb}^{\text{min}}}\xspace}
\newcommand{\chisq}{\ensuremath{\chi^2}\xspace}
\newcommand{\sqrts}{\ensuremath{\sqrt{s}}\xspace}
\newcommand{\ptmin}{\ensuremath{\pt^{\text{min}}}\xspace}

\newcommand{\Aselk}{\ensuremath{A^k_\text{sel}}\xspace}
\newcommand{\effselk}{\ensuremath{\epsilon^k_\text{sel}}\xspace}
\newcommand{\effbk}{\ensuremath{\epsilon^k_{\cPqb}}\xspace}
\newcommand{\Cbk}{\ensuremath{C^k_{\cPqb}}\xspace}

\newcommand{\Sonebk}{\ensuremath{S_{1\cPqb}^k}\xspace}
\newcommand{\Stwobk}{\ensuremath{S_{2\cPqb}^k}\xspace}
\newcommand{\Sotherbk}{\ensuremath{S_\text{other}^k}\xspace}

\newlength\cmsFigWidth
\ifthenelse{\boolean{cms@external}}{\setlength\cmsFigWidth{0.49\textwidth}}{\setlength\cmsFigWidth{0.65\textwidth}}
\ifthenelse{\boolean{cms@external}}{\providecommand{\cmsLeft}{upper\xspace}}{\providecommand{\cmsLeft}{left\xspace}}
\ifthenelse{\boolean{cms@external}}{\providecommand{\cmsRight}{lower\xspace}}{\providecommand{\cmsRight}{right\xspace}}

\ifthenelse{\boolean{cms@external}}{\providecommand{\suppMaterial}
  {the supplemental material [URL will be inserted by publisher]}}
{\providecommand{\suppMaterial}{Appendix~\ref{app:suppMat}}}

\cmsNoteHeader{TOP-19-007}
\title{Running of the top quark mass from proton-proton collisions at $\sqrts = 13\TeV$}

\date{\today}

\abstract{
  The running of the top quark mass is experimentally investigated for the first time. The mass of the top quark
  in the modified minimal subtraction (\msbar) renormalization scheme is extracted
  from a comparison of the differential top quark-antiquark (\ttbar) cross section as a function of the
  invariant mass of the \ttbar system to next-to-leading-order theoretical predictions.
  The differential cross section is determined at the parton level by means of a maximum-likelihood fit
  to distributions of final-state observables.
  The analysis is performed using \ttbar candidate events in the \emu channel
  in proton-proton collision data at a centre-of-mass energy of 13\TeV recorded
  by the CMS detector at the CERN LHC in 2016, corresponding to an integrated luminosity of 35.9\fbinv.
  The extracted running is found to be compatible
  with the scale dependence predicted by the corresponding renormalization group equation.
  In this analysis, the running is probed up to a scale of the order of 1\TeV.
}

\hypersetup{
pdfauthor={CMS Collaboration},
pdftitle={First measurement of the running of the top quark mass},
pdfsubject={CMS},
pdfkeywords={CMS, physics, top quark mass, QCD, renormalization}}

\maketitle

\section{Introduction}

\label{sec:introduction}

Beyond leading order in perturbation theory, the fundamental parameters of the quantum chromodynamics (QCD) Lagrangian, \ie the strong coupling constant \as
and the quark masses, are subject to renormalization. As a result, these parameters depend on the scale at which they are evaluated.
The evolution of \as and of the quark masses as a function of the scale, commonly referred to as ``running'',
is described by renormalization group equations (RGEs).
The running of \as was experimentally verified on a wide range of scales using jet production in electron-proton, positron-proton, electron-positron,
proton-antiproton, and proton-proton (\pp) collisions, as summarized, \eg in Refs.~\cite{Deur:2016tte,Khachatryan:2016mlc}.
To determine the running, the value of \as evaluated at an arbitrary reference scale is extracted in bins of a physical energy scale $Q$ and then converted to $\as(Q)$
using the corresponding RGE~\cite{Khachatryan:2016mlc}. The validity of this procedure lies in the fact that, in a calculation, the renormalization scale
is normally identified with the physical energy scale of the process. The same procedure can be used to determine the running of the mass of a quark.
In the modified minimal subtraction (\msbar) renormalization scheme, the dependence of a quark mass $m$ on the scale $\mu$
is described by the RGE
\begin{linenomath*}
\begin{equation}
\mu^2 \frac{\rd m(\mu)}{\rd \mu^2} = - \gamma(\as(\mu))~m(\mu),
\label{eq:RGE}
\end{equation}
\end{linenomath*}
where $\gamma(\as(\mu))$ is the mass anomalous dimension, which is known up to five-loop order in perturbative QCD~\cite{Baikov:2014qja,Luthe:2016xec}.
The solution of Eq.~\ref{eq:RGE} can be used to obtain the quark mass at any scale $\mu$ from the mass evaluated at an initial scale $\mu_0$.
The running of the \cPqb~quark mass was demonstrated~\cite{Behnke:2015qja} using data from various experiments at the CERN LEP~\cite{Abdallah:2008ac,Bambade:2005gqh,Barate:2000ab,Abbiendi:2001tw},
SLAC SLC~\cite{Brandenburg:1999nb}, and DESY HERA~\cite{Abramowicz:2014zub} colliders.
Measurements of charm quark pair production in deep inelastic scattering at the DESY HERA were used to determine the running of the charm
quark mass~\cite{Gizhko:2017fiu}. These measurements represent a powerful test of the validity of perturbative QCD.
Furthermore, RGEs can be modified by contributions from physics beyond the standard model, \eg in the context of supersymmetric theories~\cite{Mihaila:2013wma}.

This Letter describes the first experimental investigation of the running of the top quark mass, \mt, as defined in the \msbar scheme.
The running of \mt is extracted from a measurement of the differential top quark-antiquark pair production cross section,
\stt, as a function of the invariant mass of the \ttbar system, \mtt.
The differential cross section, \dstt, is determined at the parton level by means of
a maximum-likelihood fit to distributions of final-state observables using \ttbar candidate events in the \emu final state,
extending the method described in Ref.~\cite{Sirunyan:2018goh} to the case of a differential measurement.
This method allows the differential cross section to be constrained simultaneously with the systematic uncertainties.
In this analysis, the parton level is defined before radiation from the parton shower, which allows for a direct comparison with fixed-order theoretical predictions.
The measurement is performed using \pp collision data at $\sqrts = 13\TeV$ recorded by the CMS detector at the CERN LHC in 2016,
corresponding to an integrated luminosity of 35.9\fbinv.
The running mass, \mtmu, is extracted at next-to-leading order (NLO) in QCD as a function of \mtt by comparing fixed-order theoretical predictions at NLO
to the measured \dstt. The running of \mt is probed up to a scale of the order of 1\TeV.

\section{The CMS detector and Monte Carlo simulation}
\label{sec:CMS_MC}
The central feature of the CMS apparatus is a superconducting solenoid of 6\unit{m} internal diameter, providing a magnetic field of 3.8\unit{T}. Within the solenoid volume are a silicon pixel and strip tracker, a lead tungstate crystal electromagnetic calorimeter (ECAL), and a brass and scintillator hadron calorimeter, each composed of a barrel and two endcap sections. Forward calorimeters extend the pseudorapidity ($\eta$) coverage provided by the barrel and endcap detectors. Muons are detected in gas-ionization chambers embedded in the steel flux-return yoke outside the solenoid.
A two-level trigger system selects events of interest for analysis~\cite{Khachatryan:2016bia}.
A more detailed description of the CMS detector, together with a definition of the coordinate system used and the relevant kinematic variables, can be found in Ref.~\cite{Chatrchyan:2008zzk}.

The particle-flow (PF) algorithm~\cite{CMS-PRF-14-001} aims to reconstruct and identify electrons, muons, photons, charged and neutral hadrons in an event, with an optimized combination of information from the various elements of the CMS detector.
The energy of electrons is determined from a combination of the electron momentum at the primary interaction vertex as determined by the tracker, the energy of the corresponding ECAL cluster, and the energy sum of all bremsstrahlung photons spatially compatible with originating from the electron track~\cite{Khachatryan:2015hwa}. The momentum of muons is obtained from the curvature of the corresponding track~\cite{Sirunyan:2018fpa}.
Jets are reconstructed from the PF candidates using the anti-\kt clustering algorithm with a distance parameter of 0.4~\cite{Cacciari:2008gp, Cacciari:2011ma}, and the
jet momentum is determined as the vectorial sum of all particle momenta in the jet. The missing transverse momentum vector is computed as the negative vector sum of the transverse momenta (\pt) of all the PF candidates in an event. Jets originating from the hadronization of \cPqb~quarks (\cPqb~jets) are identified (\cPqb~tagged) using the combined secondary vertex~\cite{Sirunyan:2017ezt} algorithm, using a working point that corresponds to an average \cPqb~tagging efficiency of 41\% for simulated \ttbar events, and an average misidentification probability of 0.1\% and 2.2\% for light-flavour jets and \cPqc~jets, respectively~\cite{Sirunyan:2017ezt}.

In this analysis, the same Monte Carlo (MC) simulations as in Ref.~\cite{Sirunyan:2018goh} are used. In particular, \ttbar, \tW, and Drell--Yan (DY) events are
simulated using the \POWHEG~v2~\cite{Nason:2004rx,Frixione:2007vw,Alioli:2010xd,Frixione:2007nw,Re:2010bp,Alioli:2010qp} NLO MC generator interfaced to
\PYTHIA~8.202~\cite{Sjostrand:2014zea} for the modelling of the parton shower and using the CUETP8M2T4 underlying event tune~\cite{CMS-PAS-TOP-16-021,Skands:2014pea}.
In the simulation, the proton structure is described by means of the NNPDF3.0~\cite{Ball:2014uwa} parton distribution function (PDF) set.
The largest background contributions are represented by \tW and DY production.
Other background processes include \Wjets production and diboson events, while the contribution
from QCD multijet production is found to be negligible.
Contributions from all background processes are estimated from simulation and are normalized to their predicted cross section.
Further details on the MC simulation of the backgrounds can be found in Ref.~\cite{Sirunyan:2018goh}.

\section{Event selection and systematic uncertainties}
\label{sec:sel_kin}

Events are collected using a combination of triggers which require either one electron with
$\pt > 12 \GeV$ and one muon with $\pt > 23\GeV$,
or one electron with $\pt > 23 \GeV$ and one muon with $\pt > 8 \GeV$, or one electron with $\pt > 27\GeV$,
or one muon with $\pt > 24\GeV$.
In the analysis, tight isolation requirements are applied to electrons and muons based on the ratio of the scalar sum of the \pt of neighbouring
PF candidates to the \pt of the lepton candidate. Events are then required to contain at least one electron
and one muon of opposite electric charge with $\pt > 25\GeV$ for the leading
and $\pt > 20\GeV$ for the subleading lepton, and $|\eta| < 2.4$. This kinematic selection defines the visible phase space.
In events with more than two leptons, the two leptons of opposite charge with the highest \pt are used.
Jets with $\pt>30\GeV$ and $|\eta| <2.4$ are considered, but no requirement on the number of reconstructed jets or \cPqb-tagged jets is imposed.
Further details on the event selection can be found in Ref.~\cite{Sirunyan:2018goh}.

In events with at least two jets, the invariant mass of the \ttbar system is estimated by means of the
kinematic reconstruction algorithm described in Ref.~\cite{Sirunyan:2018ucr}. The reconstructed invariant mass is indicated with \mttreco.
The kinematic reconstruction algorithm examines all possible combinations of reconstructed jets and leptons
and solves a system of equations under the assumptions that
the invariant mass of the reconstructed \PW boson is 80.4\GeV
and that the missing transverse momentum originates solely from the two neutrinos coming from the leptonic decays of the \PW bosons.
In addition, the kinematic reconstruction algorithm requires an assumption on the value of the top quark mass, \mtkin.
Any possible bias due to the choice of this value is avoided by incorporating the dependence on \mtkin in the fit described
in Section~\ref{sec:fit_procedure}. To estimate this dependence, the kinematic reconstruction and the event selection
are repeated with three different choices of \mtkin,
corresponding to 169.5, 172.5, and 175.5\GeV, and the top quark mass used in the MC simulation, \mtmc, is varied accordingly.
The parameter $\mtkin = \mtmc$ is then treated as a free parameter of the fit.

The sources of systematic uncertainties are classified as experimental and modelling uncertainties.
Experimental uncertainties are related to the corrections applied to the MC simulation.
These include uncertainties associated with trigger and lepton identification efficiencies, jet
energy scale~\cite{Khachatryan:2016kdb} and resolution~\cite{CMS-PAS-JME-16-003}, lepton energy scales, \cPqb~tagging efficiencies~\cite{Sirunyan:2017ezt},
and the uncertainty in the integrated luminosity~\cite{CMS-PAS-LUM-17-001}. Modelling uncertainties are related to the simulation of the \ttbar signal,
and include matrix-element scale variations in the \POWHEG simulation~\cite{Cacciari:2003fi,Catani:2003zt}, scale variations in the parton shower~\cite{Skands:2014pea},
variations in the matching scale between the matrix element and the parton shower~\cite{CMS-PAS-TOP-16-021}, uncertainties in the underlying event tune~\cite{CMS-PAS-TOP-16-021},
the PDFs~\cite{Dulat:2015mca}, the \PB~hadron branching fraction and fragmentation function~\cite{Bowler:1981sb,Peterson:1982ak}, and
uncertainties related to the choice of the colour reconnection model~\cite{Argyropoulos:2014zoa,Christiansen:2015yqa}.
Furthermore, as in previous CMS analyses, \eg~\cite{Sirunyan:2018goh,Sirunyan:2019nlw,Sirunyan:2019lnl},
an uncertainty that accounts for the observed difference in the shape of the top quark \pt distribution between data and
simulation~\cite{Sirunyan:2018ucr,Sirunyan:2017mzl,Khachatryan:2016mnb} is applied.
The dependence on the top quark width has been investigated and was found to be negligible.
Other sources of uncertainty include the modelling of the additional \pp interactions within the same or nearby bunch crossings
and the normalization of background processes.
For the latter, an uncertainty of 30\% is assigned to the normalization of each background process.
Further details on the sources of systematic uncertainties and the considered variations can be found in Ref.~\cite{Sirunyan:2018goh}.

\begin{figure}[htbp!]
  \centering
  \includegraphics[width=0.5\textwidth]{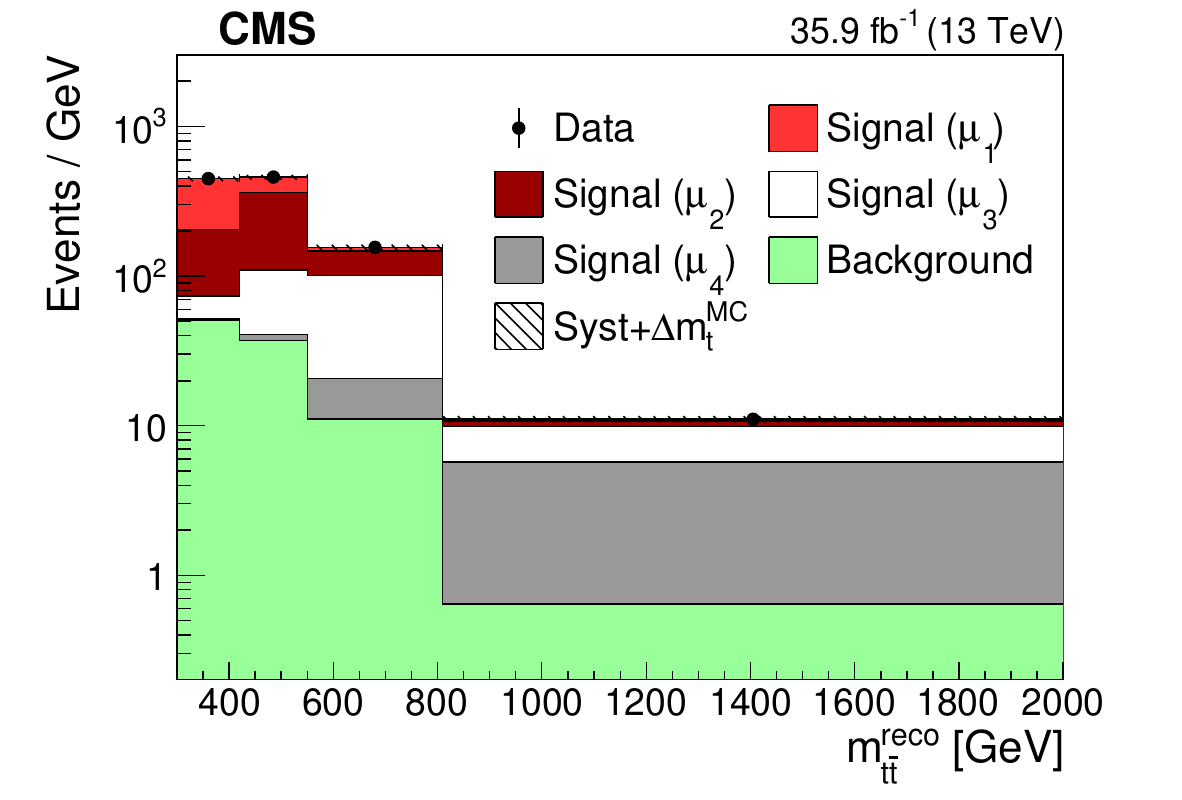}
  \caption{Distribution of \mttreco after the fit to the data, with the same binning as used in the fit.
    The hatched band corresponds to the total uncertainty in the predicted yields, including the contribution
    from \mtmc ($\Delta \mtmc$) and all correlations.
    The \ttbar MC sample is split into
    four subsamples, denoted with ``Signal~$(\muk)$'', corresponding to bins of \mtt at the parton level.
    The first and last bins contain all events with $\mttreco < 420 \GeV$ and $\mttreco > 810\GeV$, respectively.
    \label{fig:mtt_postfit}}
\end{figure}

The simulated \ttbar sample is split into four subsamples corresponding to bins of \mtt at the parton level.
Each subsample is treated as an independent signal process, representing the \ttbar production at the scale \muk,
which is chosen to be the centre-of-gravity of bin $k$, defined as the mean value of \mtt in that bin.
The subsample corresponding to the bin $k$ is denoted with ``Signal~$(\muk)$''.
The \mtt bin boundaries, the corresponding fraction of simulated events in each bin, and the representative scales \muk
are summarized in Table~\ref{tab:binning}, where the  values are estimated from the nominal \POWHEG simulation.
The width of each bin, \Dmtt, is chosen taking into account the resolution in \mttreco.
Figure~\ref{fig:mtt_postfit} shows the distribution of \mttreco
after the fit to the data, which is described in the next section.

\begin{table}[htbp!]
  \centering
  \topcaption{\label{tab:binning}
    The \mtt bin boundaries, the corresponding fraction of events in the \POWHEG simulation, and the representative scale \muk.
  }
  \begin{tabular}{cccc}
    Bin & \mtt [{\GeVns}] & Fraction [\%] & \muk [{\GeVns}] \\ \hline
    1 & $<$420      & 30 & 384 \\
    2 & 420--550      & 39 & 476 \\
    3 & 550--810      & 24 & 644 \\
    4 & $>$810 & 7  & 1024 \\
  \end{tabular}
\end{table}

\section{Fit procedure and cross section results}
\label{sec:fit_procedure}

The differential \ttbar cross section at the parton level is measured by means of a maximum-likelihood fit to
distributions of final-state observables
where the systematic uncertainties are treated as nuisance parameters.
In the likelihood, the number of events in each bin of any distribution of final-state observables is assumed to follow a Poisson distribution.
With $\sttk = (\dstt) \Dmtt$ being the total \ttbar cross section in the bin $k$ of \mtt, the expected number of events
in the bin $i$ of any of the considered final-state distributions, denoted with $\nu_i$, can be written as
\begin{linenomath*}
\begin{equation}
\nu_i = \sum_{k=1}^{4}s_i^k(\sttk,\mtmc,\vec{\lambda}) + \sum_{j} b^j_i(\mtmc,\vec{\lambda}).
\label{eq:expectev}
\end{equation}
\end{linenomath*}
Here, $s_i^k$ indicates the expected number of \ttbar events in the bin $k$ of \mtt
and depends on \sttk, \mtmc, and the nuisance parameters $\vec{\lambda}$.
Similarly, $b^j_i$ represents the expected number of background events from a source $j$ and depends
on \mtmc and the nuisance parameters $\vec{\lambda}$.
The dependence of the background processes on \mtmc is introduced not only by the contribution of \tW and semileptonic \ttbar events,
but also by the choice of \mtkin in the kinematic reconstruction.
Equation~\ref{eq:expectev}, which relates the various \sttk (and hence the parton-level differential cross section) to distributions of final-state observables,
embeds the detector response and its parametrized dependence on the systematic uncertainties.
Therefore, the maximization of the likelihood function provides results for \sttk that are automatically unfolded to the parton level.
This method (described, \eg in Ref.~\cite{Sirunyan:2018kta}) is also referred to as maximum-likelihood unfolding and,
unlike other unfolding techniques, allows the nuisance parameters to be constrained simultaneously with the differential cross section.
The unfolding problem was found to be well-conditioned, and therefore no regularization is needed.
The expected signal and background distributions contributing to the fit are modelled with templates constructed using simulated samples.

Selected events are categorized according to the number of \cPqb-tagged jets, as
events with 1~\cPqb-tagged jet, 2~\cPqb-tagged jets, or a different number of \cPqb-tagged jets (zero or more than two).
The effect of the systematic uncertainties on the normalization of the different signals in each of these categories is
parametrized using multinomial probabilities. In particular, based on the \ttbar topology, the number of events with one (\Sonebk), two (\Stwobk),
or a different number of \cPqb-tagged jets (\Sotherbk) in each bin of \mtt is expressed as:
\begin{linenomath}
  \begin{eqnarray}
      \Sonebk  &=& \intlumi \sttk \Aselk \effselk 2\effbk (1-\Cbk \effbk) , \label{eq:S1} \\
      \Stwobk  &=& \intlumi \sttk \Aselk \effselk \Cbk (\effbk)^2 , \label{eq:S2} \\
      \Sotherbk &=& \intlumi \sttk \Aselk \effselk \left[1-2\effbk(1-\Cbk \effbk)-\Cbk(\effbk)^2\right] \label{eq:Sother} .
  \end{eqnarray}
\end{linenomath}
Here, \intlumi is the integrated luminosity, \Aselk is the acceptance of the event selection in the \mtt bin $k$,
and \effselk represents the efficiency for an event in the visible phase space to pass the full event selection.
The acceptance \Aselk is defined as  the fraction of \ttbar events in the bin $k$ that, at the generator (particle) level,
enter the visible phase space described in Section~\ref{sec:sel_kin}, while \effselk includes
experimental selection criteria, \eg isolation and trigger requirements.
Furthermore, \effbk represents the \cPqb~tagging probability and the parameter \Cbk accounts
for any residual correlation between the tagging of two \cPqb~jets in a \ttbar event.
The quantities \Aselk, \effselk, \effbk, and \Cbk are
determined from the signal simulation and, although they are not free parameters of the fit,
they vary according to the parameters $\vec{\lambda}$ and \mtmc. In each category, the remaining effects of the systematic uncertainties
on signal processes are treated as shape uncertainties.
The quantities $s_i^k$ in Eq.~\ref{eq:expectev} are then derived from the signal shape and normalization in the corresponding category.
In this way, a precise parametrization of the dependence of signal normalizations on the nuisance parameters and \mtmc is obtained.
In fact, the parameters in Eqs.~\ref{eq:S1}--\ref{eq:Sother} are less subject to statistical fluctuations than the $s_i^k$.

In order to constrain each individual \sttk, events with at least two jets are further divided into subcategories of \mttreco,
using the same binning as for \mtt (Table~\ref{tab:binning}).
The choice of the input distributions to the fit in the different event categories is summarized in Table~\ref{tab:inputs}.
The total number of events is chosen as input to the fit for all subcategories
with zero or more than two \cPqb-tagged jets, where the contribution of the background processes is the largest,
in order to mitigate the sensitivity of the measurement to the shape of the distributions of background processes.
The same choice is made for the subcategories corresponding to the last bin in \mttreco, where the statistical uncertainty
in both data and simulation is large, and for events with less than two jets, where the kinematic reconstruction cannot be performed.
In the remaining subcategories with one \cPqb-tagged jet, the minimum invariant mass found when combining the reconstructed \cPqb~jet and a lepton,
referred to as the \mlb distribution, is fitted. This distribution provides the sensitivity to constrain \mtmc~\cite{Biswas:2010sa}.
In the remaining subcategories with two \cPqb-tagged jets, the \pt spectrum of the softest selected jet in the event is used
to constrain jet energy scale uncertainties at small values of \pt, the kinematic range where systematic uncertainties are the largest.
The distributions used in the fit are compared to the data after the fit in \suppMaterial.

\begin{table}[htbp!]
  \centering
  \topcaption{\label{tab:inputs}
    Input distributions to the fit in the different event categories. The number of jets, the number of \cPqb-tagged jets,
    the number of events, and the \pt of the softest jet are denoted with \Njets, \Nb, \Nev, and ``jet \ptmin'', respectively,
    while the category corresponding to the bin $k$ in \mttreco is indicated with ``\mttreco$k$''.
  }
  \begin{tabular}{l|ccc}
    & $\Nb=1$ & $\Nb=2$ & Other \Nb \\ \hline
    $\Njets < 2$ & \Nev  & n.a. & \Nev \\
    \mttreco1 & \mlb & jet \ptmin & \Nev \\
    \mttreco2 & \mlb & jet \ptmin & \Nev \\
    \mttreco3 & \mlb & jet \ptmin & \Nev \\
    \mttreco4 & \Nev & \Nev & \Nev \\
  \end{tabular}
\end{table}

The efficiencies of the kinematic reconstruction in data and simulation have been investigated in Ref.~\cite{Sirunyan:2018ucr}
and they were found to differ by 0.2\%.
Therefore, the efficiency in the simulation is corrected to match the one in data.
An uncertainty of 0.2\% is assigned to each bin of \mtt independently.
The same uncertainty is also assigned to \ttbar events with one or two \cPqb-tagged jets, independently.
For \ttbar events with zero or more than two \cPqb-tagged jets, where the combinatorial background is larger, an uncertainty of 0.5\% is conservatively assigned.
These uncertainties are treated as uncorrelated to account for possible differences between the different \mtt bins and
categories of \cPqb-tagged jet multiplicity.
Similarly, an additional uncertainty of 1\% is assigned to the sum of the background processes, independently for
each bin of \mttreco, in order to reduce the correlation between the signal and the background templates.
The impact of these uncertainties on the final results is found to be small compared to the total uncertainty.

The dependence of the signal shapes, of the parameters \Aselk, \effselk, \effbk, and \Cbk,
and of the background contributions on \mtmc and on the nuisance parameters $\vec{\lambda}$
is modelled using second-order polynomials~\cite{Sirunyan:2018goh}.
In the fit, Gaussian priors are assumed for all the nuisance parameters.
The negative log-likelihood is then minimized, using the \Minuit program~\cite{James:1975dr},
with respect to \sttk, \mtmc, and $\vec{\lambda}$.
Finally, the fit uncertainties in the various \sttk are determined using \Minos~\cite{James:1975dr}.
Additional extrapolation uncertainties, which reflect the impact of modelling uncertainties on \Aselk,
are estimated without taking into account the constraints obtained in the visible phase space~\cite{Sirunyan:2018goh}.
Moreover, an additional uncertainty arising from the limited statistical precision of the simulation
is estimated using MC pseudo-experiments~\cite{Sirunyan:2018goh}, where templates are varied within their statistical uncertainties
taking into account the correlations between the nominal templates and the templates corresponding to the systematic variations.
The template dependencies are then rederived and the fit to the data is repeated more than ten thousand times.
For each parameter of interest, the root-mean-square of the best fit values obtained with this procedure
is taken as an additional uncertainty and added in quadrature to the total uncertainty from the fit.

\begin{figure}[htbp!]
  \centering
  \includegraphics[width=0.5\textwidth]{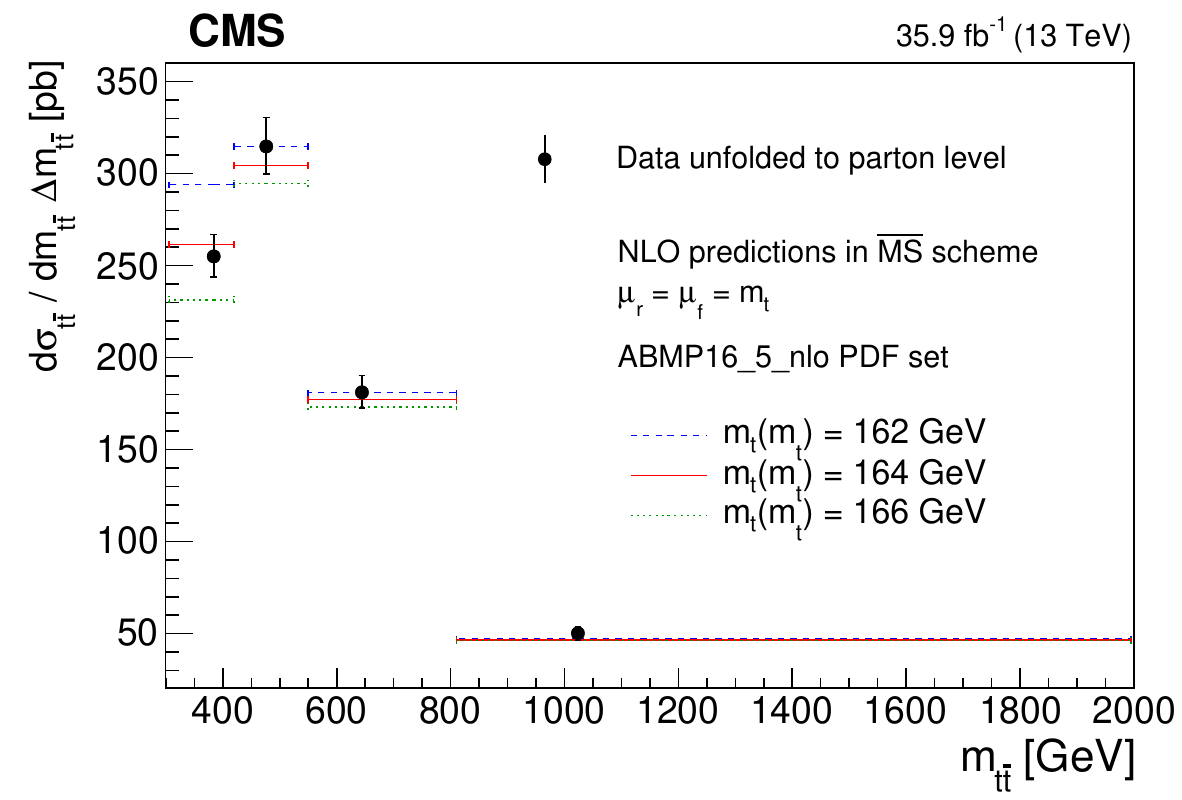}
  \caption{
    Measured values of \sttk (markers) and their uncertainties (vertical error bars)
    compared to NLO predictions in the \msbar scheme obtained with different values of \mtmt (horizontal lines of different styles).
    The values of \sttk are shown at the representative scale of the process \muk, defined as the centre-of-gravity of bin $k$ in \mtt.
    The first and last bins contain all events with $\mtt < 420 \GeV$ and $\mtt > 810\GeV$, respectively.
    \label{fig:xsec_result}}
\end{figure}

The measured \sttk are shown in Fig.~\ref{fig:xsec_result} and compared to fixed-order theoretical predictions in the \msbar scheme at NLO~\cite{Dowling:2013baa}
implemented for the purpose of this analysis in the \MCFM~v6.8 program~\cite{Campbell:2010ff,Campbell:2012uf}. In the calculation,
the renormalization scale, \mur, and factorization scale, \muf, are both set to \mt.
The \msbar mass of the top quark evaluated at the scale $\mu=\mt$ is denoted with \mtmt.
The calculation is interfaced with the ABMP16\_5\_nlo PDF set~\cite{Alekhin:2018pai}, which is the only available PDF set where \mt is treated in the \msbar scheme
and where the correlations between the gluon PDF, \as, and \mt are taken into account.
In the calculation, the value of \as at the \cPZ boson mass, \asmz, is set to the value determined in the ABMP16\_5\_nlo fit,
which in the central PDF corresponds to $0.1191$~\cite{Alekhin:2018pai}.
In order to demonstrate the sensitivity to the top quark mass, predictions for \dstt obtained with different values of \mtmt are shown.
Furthermore, it is worth noting that this method provides a cross section result with significantly improved precision
compared to measurements that perform unfolding as a separate step, \eg as the one described in Ref.~\cite{Sirunyan:2018ucr}.

The dominant uncertainties in the measured \sttk are associated with the integrated luminosity, the lepton identification efficiencies,
the jet energy scales and, at large \mtt, the modelling of the top quark \pt. The two latter uncertainties are marginally constrained in the fit,
while the first two are not constrained.
Furthermore, the post-fit values of all nuisance parameters are found to be compatible with their pre-fit value, within one standard deviation.
The numerical values of the measured \sttk, their correlations, the impact of the various sources of uncertainty,
and the pulls and constraints of the nuisance parameters related to the modelling uncertainties can be found in \suppMaterial.

\section{Extraction of the running of the top quark mass}
\label{sec:running_extraction}

The measured differential cross section is used to extract the running of the top quark \msbar mass at NLO as a function of the scale $\mu = \mtt$.
The procedure is similar to the one used to extract the running of the charm quark mass~\cite{Gizhko:2017fiu}.
The value of \mtmt is determined independently in each bin of \mtt
from a \chisq fit of fixed-order theoretical predictions at NLO to the measured \sttk.
The theoretical predictions are obtained as described in Section~\ref{sec:fit_procedure} for Fig.~\ref{fig:xsec_result}.
The \chisq definition follows the one described in Ref.~\cite{Barlow:2004wj}, which accounts for asymmetries in
the input uncertainties.
The extracted \mtmt are then converted to \mtmuk using the \RunDec~v3.0 program~\cite{Schmidt:2012az},
where \muk is the representative scale of the process in a given bin of \mtt, as described in Section~\ref{sec:sel_kin}.
As relevant in a NLO calculation, the conversion is performed with one-loop precision, assuming five active flavours ($\nf = 5$) and $\asmz = 0.1191$
consistently with the used PDF set.
This procedure is equivalent to extracting directly \mtmuk in each bin.
Furthermore, the result does not depend on the exact choice of \muk, provided that it is representative of the physical energy scale of the process.
In fact, a change in \muk would correspond to a change in \mtmuk according to the RGE.
The extracted values of \mtmuk and their uncertainties can be found in \suppMaterial.

In order to benefit from the cancellation of correlated uncertainties in the measured \sttk,
the ratios of the various \mtmuk to \mtmutwo are considered.
In particular, the quantities $r_{12} = \mt(\mu_1)/\mtmutwo$, $r_{32} = \mt(\mu_3)/\mtmutwo$, and $r_{42} = \mt(\mu_4)/\mtmutwo$ are extracted.
With this approach the running of \mt, \ie the quantity predicted by the RGE (Eq.~\ref{eq:RGE}), is accessed directly.
The measurement at the scale $\mu_2$ is chosen as a reference in order to minimize the correlation between the extracted ratios.

Four different types of systematic uncertainty are considered for the ratios: the uncertainty in the various \sttk in the visible phase space (referred to as fit uncertainty),
the extrapolation uncertainties, the uncertainties in the proton PDFs, and the uncertainty in the value of \asmz.
The fit uncertainty includes experimental and modelling uncertainties described in Section~\ref{sec:sel_kin}.
Scale variations in the \MCFM predictions are not performed,
since the scale dependence of \mt is being investigated at a fixed order in perturbation theory.
In fact, scale variations in the hard scattering cross section are conventionally performed as a means of estimating
the effect of missing higher order corrections and are therefore not applicable in this context.

Uncertainties in the proton PDFs affect the \MCFM prediction and therefore the extracted values of the various \mtmuk. In order to estimate
their impact, the calculation is repeated for each eigenvector of the PDF set and the differences in the extracted ratios are added
in quadrature to yield the total PDF uncertainties. In the ABMP16\_5\_nlo PDF set, \asmz is determined simultaneously with the PDFs,
therefore its uncertainty is incorporated in that of the PDFs. However, the uncertainty in \asmz also affects the \RunDec conversion from \mtmt to \mtmuk.
This effect is estimated independently and is found to be negligible.

The impact of extrapolation uncertainties is estimated by varying the measured \sttk within their extrapolation uncertainty, separately for
each source and simultaneously in the different bins in \mtt, taking the correlations into account.
The various contributions are added in quadrature to yield the total extrapolation uncertainty.

The correlations between the extracted masses arising from the fit uncertainty are estimated using MC pseudo-experiments,
taking the correlations between the measured \sttk as inputs.
The uncertainties are then propagated to the ratios using linear uncertainty propagation, taking the estimated correlations into account.
The numerical values of the ratios are determined to be:
\begin{linenomath}
  \begin{equation*}
    \begin{aligned}
      r_{12} &= 1.030 \pm 0.018~(\text{fit}) ~^{+0.003}_{-0.006}~(\text{PDF+\as}) ~^{+0.003}_{-0.002}~(\text{extr})  ,  \\
      r_{32} &= 0.982 \pm 0.025~(\text{fit}) ~^{+0.006}_{-0.005}~(\text{PDF+\as}) \pm 0.004~(\text{extr})  ,  \\
      r_{42} &= 0.904 \pm 0.050~(\text{fit}) ~^{+0.019}_{-0.017}~(\text{PDF+\as}) ~^{+0.017}_{-0.013}~(\text{extr})  .
    \end{aligned}
  \end{equation*}
\end{linenomath}
Here, the fit uncertainty (fit), the combination of PDF and \as uncertainty (PDF+\as), and the extrapolation uncertainty (extr) are given.
The most relevant sources of experimental uncertainty are the integrated luminosity,
the lepton identification efficiencies, and the jet energy scale and resolution.
Among modelling uncertainties related to the \POWHEG+\PYTHIA~8 simulation of the \ttbar signal, the largest contributions originate from
the scale variations in the parton shower, the uncertainty in the shape of the \pt spectrum of the top quark,
and the matching scale between the matrix element and the parton shower. The statistical uncertainties are found to be negligible.
The correlations between the ratios arising from the fit uncertainty are investigated using a pseudo-experiment procedure which consists in repeating
the extraction of the ratios using pseudo-measurements of \sttk, generated according to the corresponding fitted values,
uncertainties, and correlations.
With \rhoik being the correlation between $r_{i2}$ and $r_{k2}$, the results are $\rho_{13} = \ratiocorronetwo$,
$\rho_{14} = \ratiocorronethree$, and $\rho_{34} = \ratiocorrtwothree$.

\begin{figure}[htb!]
  \centering
  \includegraphics[width=0.495\textwidth]{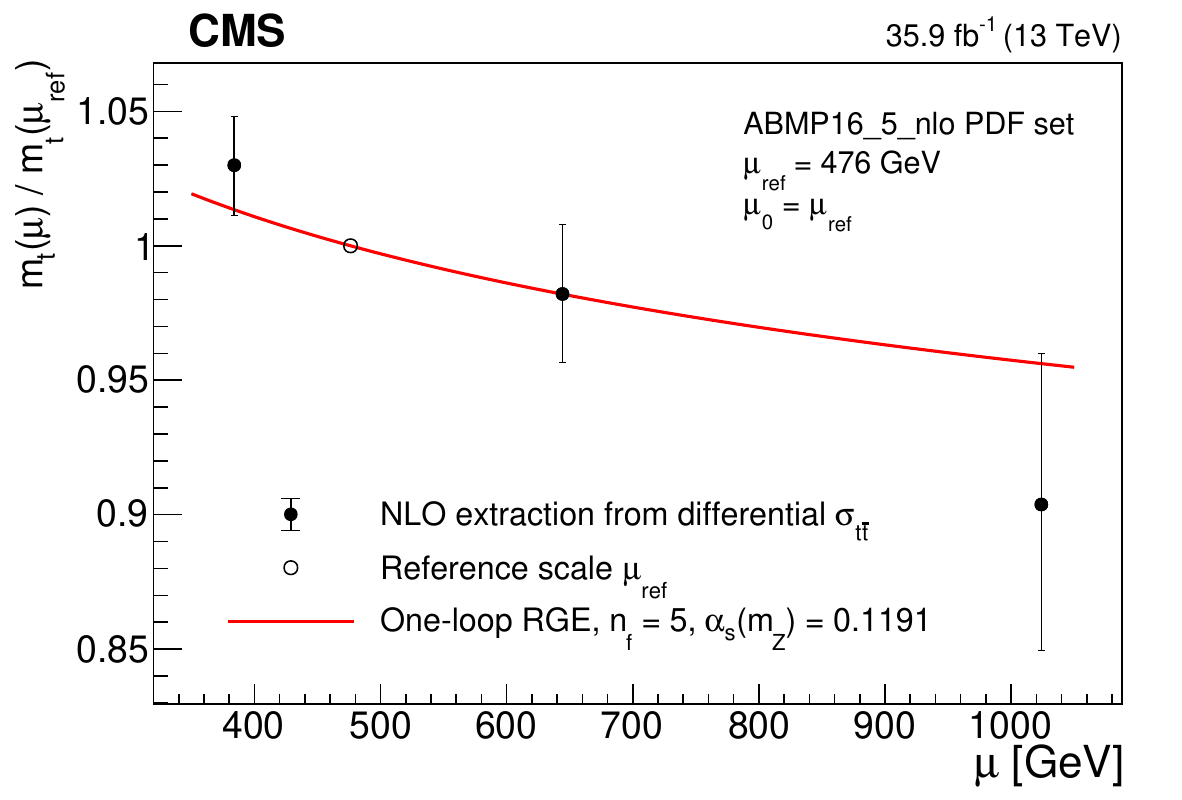}
  \includegraphics[width=0.495\textwidth]{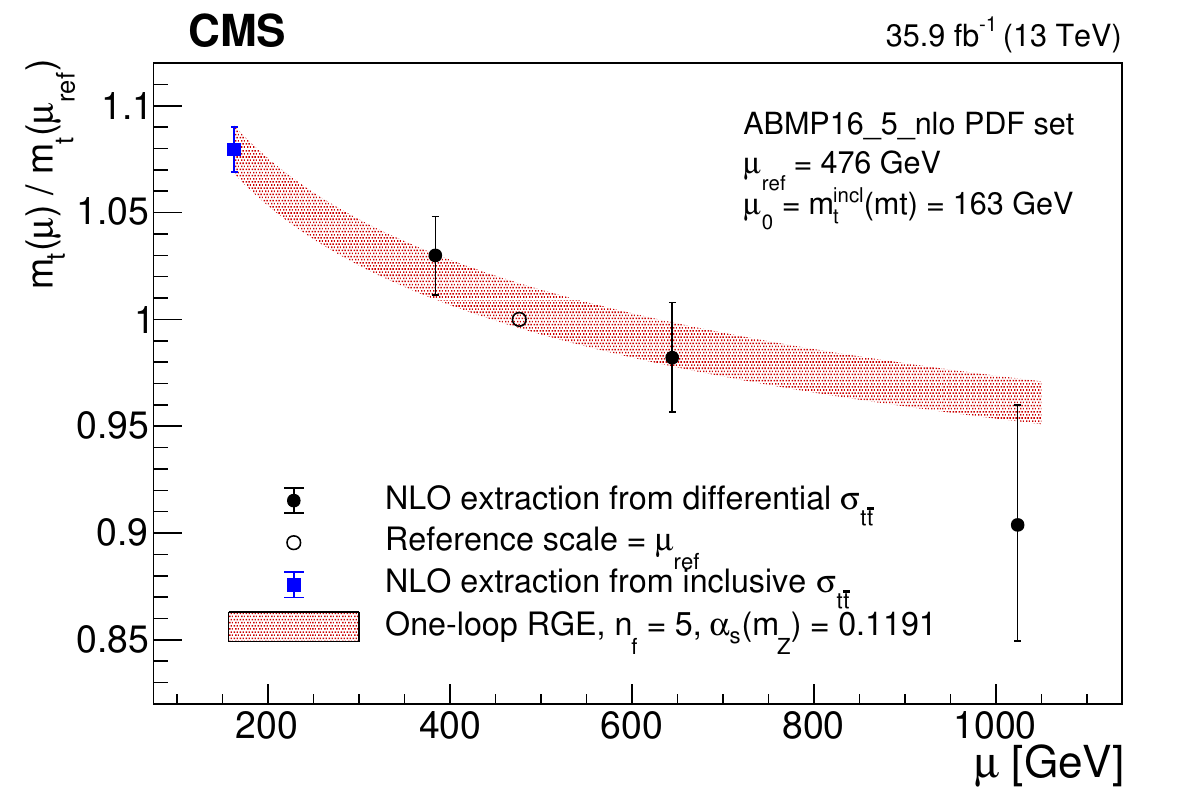}
  \caption{
    Extracted running of the top quark mass $\mtmu/\mtmuref$ compared to the RGE prediction at one-loop precision, with $\nf = 5$,
    evolved from the initial scale $\mu_0 = \muref = 476\GeV$ (\cmsLeft). The result is compared to the value of $\mtot/\mtmuref$,
    where \mtot is the value of \mtmt extracted from the inclusive cross section measured in Ref.~\cite{Sirunyan:2018goh}, which is based on the same data set.
    The uncertainty in \mtot is evolved from the initial scale $\mu_0 = \mtot$, which corresponds to about 163\GeV, using the same RGE prediction (\cmsRight).
    \label{fig:ratios_mtmu}}
\end{figure}

The extracted ratios $\mtmuk/\mtmutwo$ are shown in Fig.~\ref{fig:ratios_mtmu} (\cmsLeft) together with
the RGE prediction (Eq.~\ref{eq:RGE}) at one-loop precision.
In the figure, the reference scale $\mu_2$ is indicated with \muref, and the RGE evolution is calculated from the initial scale $\mu_0 = \muref$.
Good agreement between the extracted running and the RGE prediction is observed.

For comparison, the \msbar mass of the top quark is also extracted from the inclusive cross section measured in Ref.~\cite{Sirunyan:2018goh},
using \Hathor~2.0~\cite{Aliev:2010zk} predictions at NLO interfaced with the ABMP16\_5\_nlo PDF set, and is denoted with \mtot.
Fig.~\ref{fig:ratios_mtmu} (\cmsRight) compares the extracted ratios $\mtmuk/\mtmutwo$
to the value of $\mtot/\mtmutwo$. The uncertainty in \mtot includes fit, extrapolation, and PDF uncertainties,
and is evolved to higher scales, while the value of \mtmutwo
in the ratio $\mtot/\mtmutwo$ is taken without uncertainty. Here, the RGE evolution is calculated from the initial scale $\mu_0 = \mtot$,
which corresponds to about 163\GeV.
The extracted value of \mtot and its uncertainty can be found in \suppMaterial.

Finally, the extracted running is parametrized with the function
\begin{linenomath*}
\begin{equation}
  f(x,\mu) = x\left[ \rmu -1 \right] +1 ,
  \label{eq:fit_sign}
\end{equation}
\end{linenomath*}
where $\rmu = \mtmu/\mtmutwo$ corresponds to the RGE prediction shown in Fig.~\ref{fig:ratios_mtmu} (\cmsLeft).
In particular, $f(x,\mu)$ corresponds to \rmu for $x=1$ and to 1, \ie no running, for $x=0$.
The best fit value for $x$, denoted with \xmin, is determined via a \chisq fit to the extracted ratios
taking the correlations \rhoik into account, and is found to be
\begin{linenomath*}
  \begin{equation*}
  \xmin =  2.05 \pm 0.61~ (\text{fit}) ~^{+0.31}_{-0.55}~ (\text{PDF}+\as) ~ ^{+0.24}_{-0.49}~ (\text{extr}).
\end{equation*}
\end{linenomath*}
The result shows agreement between the extracted running and the RGE prediction at one-loop precision within \agreement standard deviations in the Gaussian approximation
and excludes the no-running hypothesis at above 95\% confidence level (2.1 standard deviations) in the same approximation.

\section{Summary}
\label{sec:summary}

In this Letter, the first experimental investigation of the running of the top quark mass, \mt, is presented.
The running is extracted from a measurement of the differential top quark-antiquark (\ttbar) cross section
as a function of the invariant mass of the \ttbar system, \mtt.
The differential \ttbar cross section, \dstt, is determined at the parton level
using a maximum-likelihood fit to distributions of final-state observables, using \ttbar candidate events in the \emu channel.
This technique allows the nuisance parameters to be constrained simultaneously
with the differential cross section in the visible phase space and therefore provides results with significantly improved precision
compared to conventional procedures in which the unfolding is performed as a separate step.
The analysis is performed using proton-proton collision data at a centre-of-mass energy
of 13\TeV recorded by the CMS detector at the CERN LHC in 2016, corresponding to an integrated luminosity of 35.9\fbinv.

The running mass \mtmu, as defined in the modified minimal subtraction (\msbar) renormalization scheme, is extracted at one-loop precision
as a function of \mtt by comparing fixed-order theoretical predictions at next-to-leading order to the measured \dstt.
The extracted running of \mt is found to be in agreement with the prediction of the corresponding renormalization group equation,
within \agreement standard deviations, and the no-running hypothesis is excluded at above 95\% confidence level.
The running of \mt is probed up to a scale of the order of 1\TeV.

\begin{acknowledgments}
We congratulate our colleagues in the CERN accelerator departments for the excellent performance of the LHC and thank the technical and administrative staffs at CERN and at other CMS institutes for their contributions to the success of the CMS effort. In addition, we gratefully acknowledge the computing centres and personnel of the Worldwide LHC Computing Grid for delivering so effectively the computing infrastructure essential to our analyses. Finally, we acknowledge the enduring support for the construction and operation of the LHC and the CMS detector provided by the following funding agencies: BMBWF and FWF (Austria); FNRS and FWO (Belgium); CNPq, CAPES, FAPERJ, FAPERGS, and FAPESP (Brazil); MES (Bulgaria); CERN; CAS, MoST, and NSFC (China); COLCIENCIAS (Colombia); MSES and CSF (Croatia); RPF (Cyprus); SENESCYT (Ecuador); MoER, ERC IUT, PUT and ERDF (Estonia); Academy of Finland, MEC, and HIP (Finland); CEA and CNRS/IN2P3 (France); BMBF, DFG, and HGF (Germany); GSRT (Greece); NKFIA (Hungary); DAE and DST (India); IPM (Iran); SFI (Ireland); INFN (Italy); MSIP and NRF (Republic of Korea); MES (Latvia); LAS (Lithuania); MOE and UM (Malaysia); BUAP, CINVESTAV, CONACYT, LNS, SEP, and UASLP-FAI (Mexico); MOS (Montenegro); MBIE (New Zealand); PAEC (Pakistan); MSHE and NSC (Poland); FCT (Portugal); JINR (Dubna); MON, RosAtom, RAS, RFBR, and NRC KI (Russia); MESTD (Serbia); SEIDI, CPAN, PCTI, and FEDER (Spain); MOSTR (Sri Lanka); Swiss Funding Agencies (Switzerland); MST (Taipei); ThEPCenter, IPST, STAR, and NSTDA (Thailand); TUBITAK and TAEK (Turkey); NASU and SFFR (Ukraine); STFC (United Kingdom); DOE and NSF (USA).

\hyphenation{Rachada-pisek} Individuals have received support from the Marie-Curie programme and the European Research Council and Horizon 2020 Grant, contract Nos.\ 675440, 752730, and 765710 (European Union); the Leventis Foundation; the A.P.\ Sloan Foundation; the Alexander von Humboldt Foundation; the Belgian Federal Science Policy Office; the Fonds pour la Formation \`a la Recherche dans l'Industrie et dans l'Agriculture (FRIA-Belgium); the Agentschap voor Innovatie door Wetenschap en Technologie (IWT-Belgium); the F.R.S.-FNRS and FWO (Belgium) under the ``Excellence of Science -- EOS" -- be.h project n.\ 30820817; the Beijing Municipal Science \& Technology Commission, No. Z181100004218003; the Ministry of Education, Youth and Sports (MEYS) of the Czech Republic; the Lend\"ulet (``Momentum") Programme and the J\'anos Bolyai Research Scholarship of the Hungarian Academy of Sciences, the New National Excellence Program \'UNKP, the NKFIA research grants 123842, 123959, 124845, 124850, 125105, 128713, 128786, and 129058 (Hungary); the Council of Science and Industrial Research, India; the HOMING PLUS programme of the Foundation for Polish Science, cofinanced from European Union, Regional Development Fund, the Mobility Plus programme of the Ministry of Science and Higher Education, the National Science Center (Poland), contracts Harmonia 2014/14/M/ST2/00428, Opus 2014/13/B/ST2/02543, 2014/15/B/ST2/03998, and 2015/19/B/ST2/02861, Sonata-bis 2012/07/E/ST2/01406; the National Priorities Research Program by Qatar National Research Fund; the Ministry of Science and Education, grant no. 3.2989.2017 (Russia); the Programa Estatal de Fomento de la Investigaci{\'o}n Cient{\'i}fica y T{\'e}cnica de Excelencia Mar\'{\i}a de Maeztu, grant MDM-2015-0509 and the Programa Severo Ochoa del Principado de Asturias; the Thalis and Aristeia programmes cofinanced by EU-ESF and the Greek NSRF; the Rachadapisek Sompot Fund for Postdoctoral Fellowship, Chulalongkorn University and the Chulalongkorn Academic into Its 2nd Century Project Advancement Project (Thailand); the Nvidia Corporation; the Welch Foundation, contract C-1845; and the Weston Havens Foundation (USA).
\end{acknowledgments}

\bibliography{auto_generated}

\ifthenelse{\boolean{cms@external}}{}{
  \clearpage
  \appendix
  \numberwithin{table}{section}
  \numberwithin{figure}{section}
  \section{Supplemental information \label{app:suppMat}}
  
Additional information that complements the content of the Letter is provided.
Several acronyms are used. These are: matrix element~(ME),
final state radiation~(FSR), initial state radiation~(ISR), parton shower~(PS), colour reconnection~(CR),
early resonance decay~(ERD), underlying event~(UE), branching fraction~(BF), and identification~(ID).

The distributions used in the fit are compared to the data after the fit in Figure~\ref{fig:postfit},
and the normalized fit pulls and constraints on the nuisance parameters related to the modelling uncertainties are shown in Figure~\ref{fig:pulls}.
By means of the likelihood fit, the quantities \sttk are determined to be:
\begin{align*}
  \stt^{(\mu_1)} &= 255 \pm 11 \syst \pm 2 \stat \unit{pb}, \\
  \stt^{(\mu_2)} &= 315 \pm 15 \syst \pm 2 \stat \unit{pb}, \\
  \stt^{(\mu_3)} &= 181 \pm 9 \syst \pm 1 \stat \unit{pb}, \\
  \stt^{(\mu_4)} &= 50 \pm 3 \syst \pm 1 \stat \unit{pb}. \\
\end{align*}
The correlations between the measured \sttk are given in Table~\ref{tab:corr}, while the contribution of the various sources of systematic uncertainties
to the total uncertainty can be found in Tables~\ref{tab:stt1} to~\ref{tab:stt4}. Finally, the extracted values of $\mt(\mu_k)$ are shown in Figure~\ref{fig:mtmuk},
together with the value of $\mt(\mt)$ extracted from the inclusive \ttbar cross section, as described in the Letter.
The numerical values of the extracted masses are:
\begin{linenomath}
  \begin{equation*}
    \begin{aligned}
      \mt(\mu_1) &= 155.4 \pm 0.8~(\text{fit}) \pm 0.2 ~(\text{PDF+\as}) \pm 0.1 ~(\text{extr}) ~^{+0.9}_{-0.6} ~(\text{scale})  ,  \\
      \mt(\mu_2) &= 150.9 \pm 3.0~(\text{fit}) ~^{+1.1}_{-0.7} ~(\text{PDF+\as}) ~^{+0.4}_{-0.5} ~(\text{extr}) ~^{+3.9}_{-4.3} ~(\text{scale})  ,  \\
      \mt(\mu_3) &= 148.2 \pm 4.6~(\text{fit}) ~^{+2.0}_{-1.4} ~(\text{PDF+\as}) ~^{+0.9}_{-1.0} ~(\text{extr}) ~^{+7.3}_{-9.5} ~(\text{scale})  ,  \\
      \mt(\mu_4) &= 136.4 \pm 9.0~(\text{fit}) ~^{+3.8}_{-3.0} ~(\text{PDF+\as}) ~^{+2.8}_{-2.3} ~(\text{extr}) ~^{+9.6}_{-16.1} ~(\text{scale})  .  \\ 
    \end{aligned}
  \end{equation*}
\end{linenomath}
The scale uncertainties are obtained by varying the renormalization and factorization scales by a factor of two,
avoiding cases in which $\mur / \muf = 1/4$ or 4. The total scale uncertainty is then defined as the envelope
of the individual variations. Similarly, the value $\mt(\mt)$ determined from the inclusive cross section is:
  \begin{equation*}
      \mt(\mt) = 162.9 \pm 1.6~(\text{fit+extr+PDF+\as}) ~^{+2.5}_{-3.0} ~(\text{scale})  .  \\
  \end{equation*}
As explained in the Letter, scale uncertainties are not considered in the extraction of the running,
which is investigated at a fixed order in perturbation theory.

\begin{figure*}[htbp]
  \centering
    \includegraphics[width=0.325\textwidth]{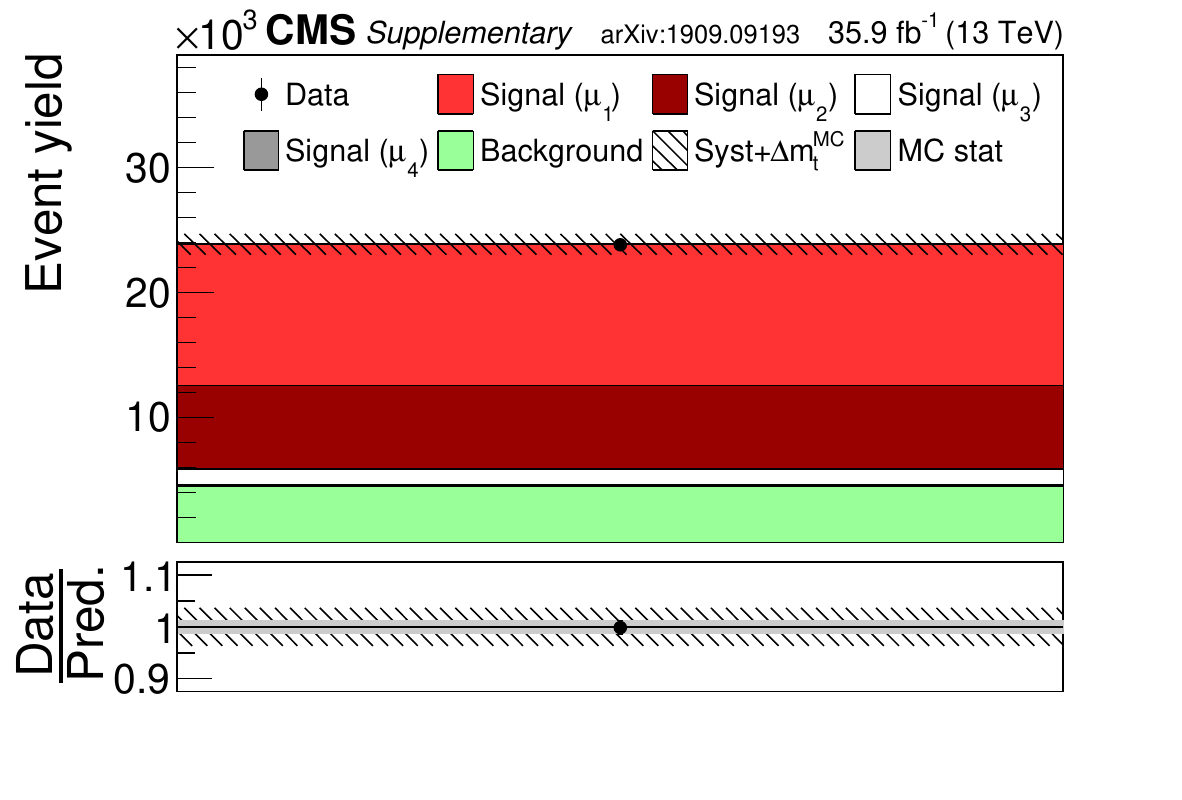}
    \includegraphics[width=0.325\textwidth]{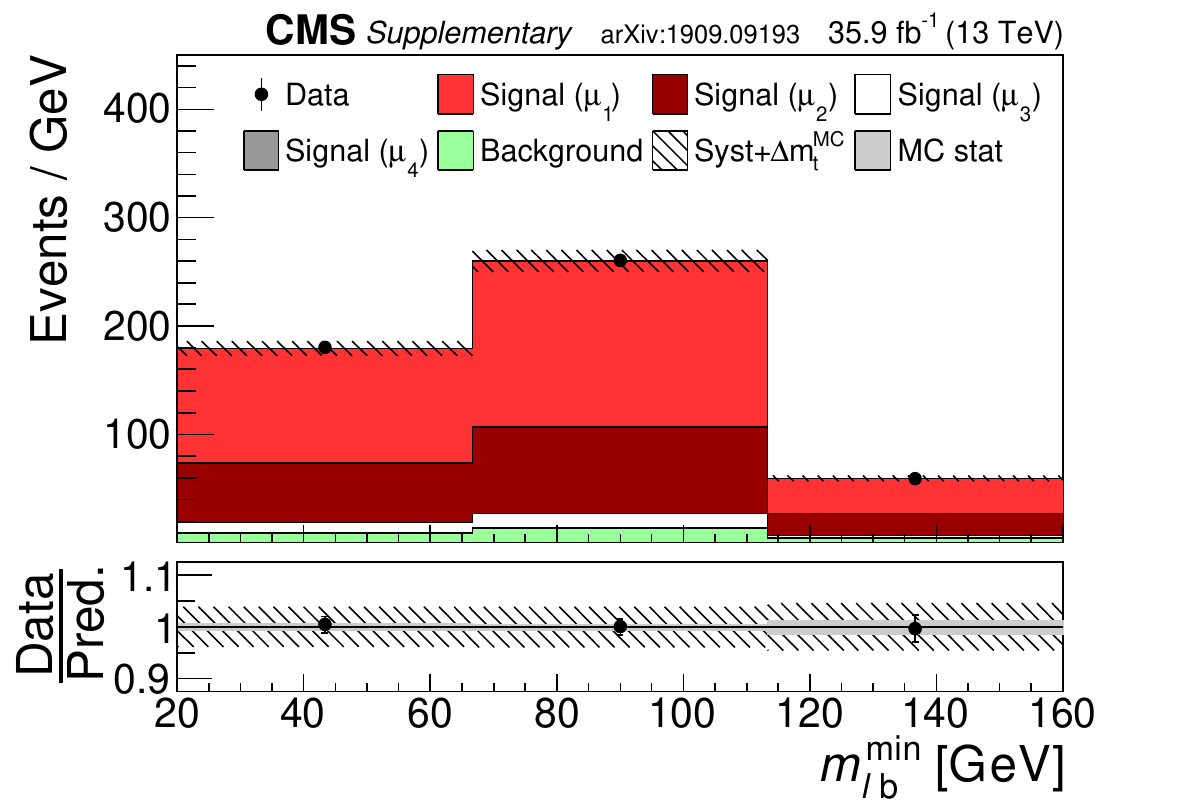}
    \includegraphics[width=0.325\textwidth]{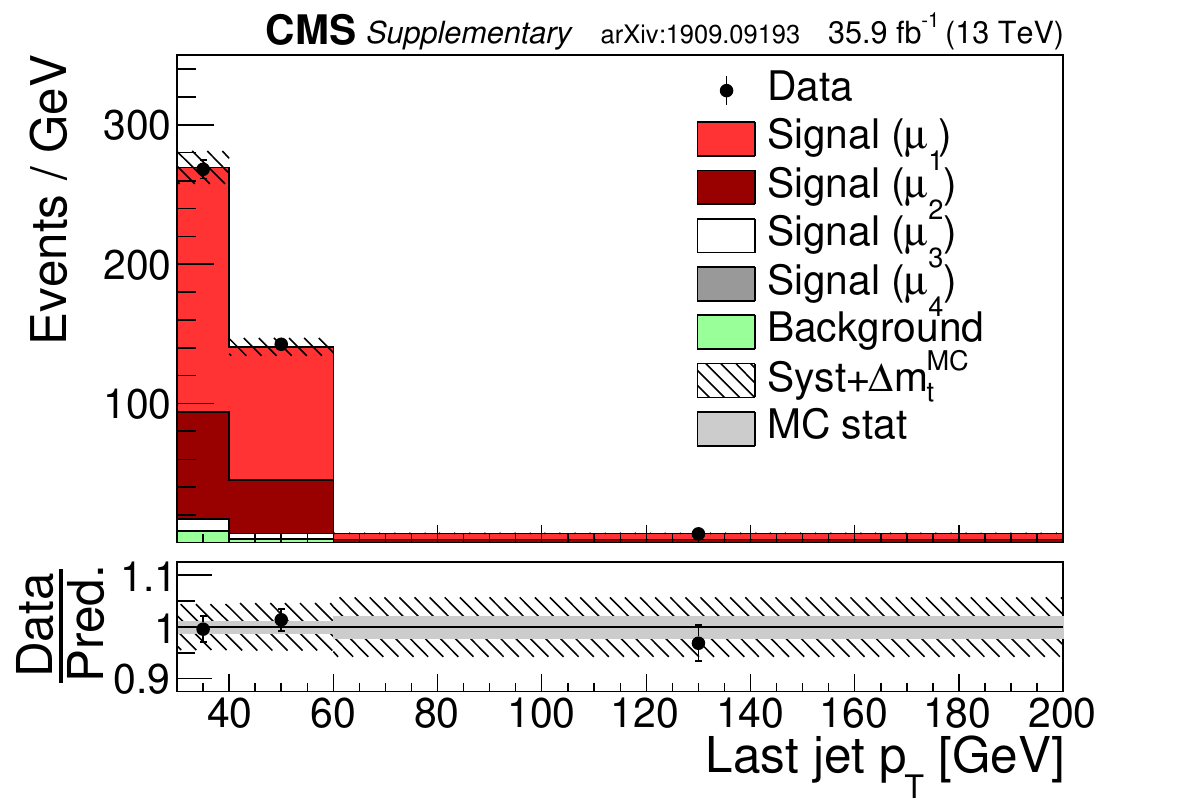}

    \includegraphics[width=0.325\textwidth]{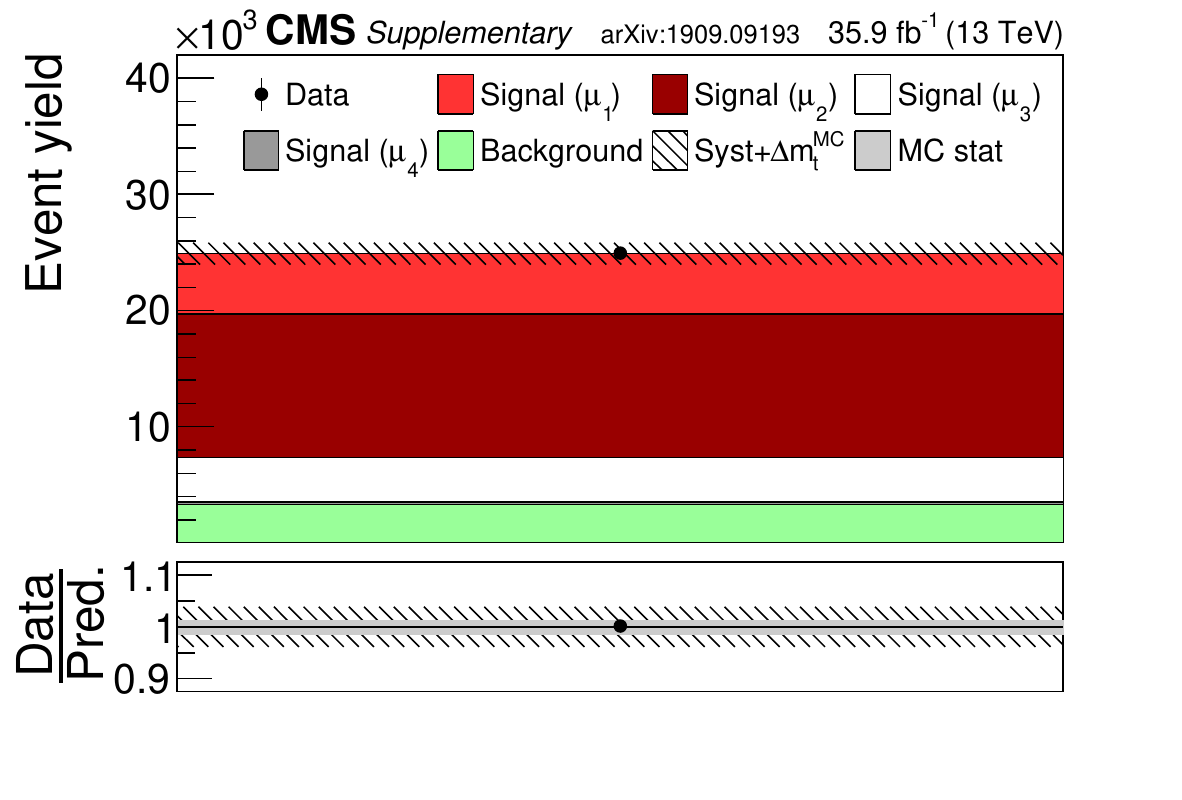}
    \includegraphics[width=0.325\textwidth]{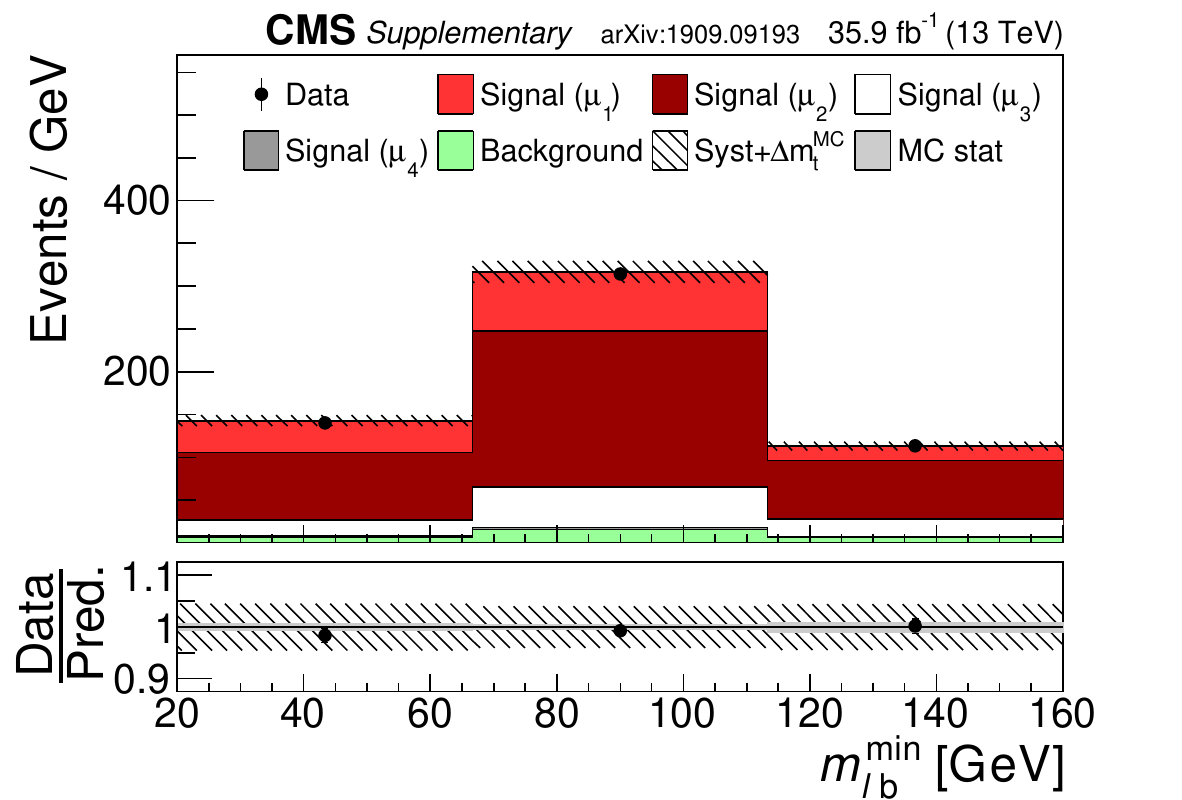}
    \includegraphics[width=0.325\textwidth]{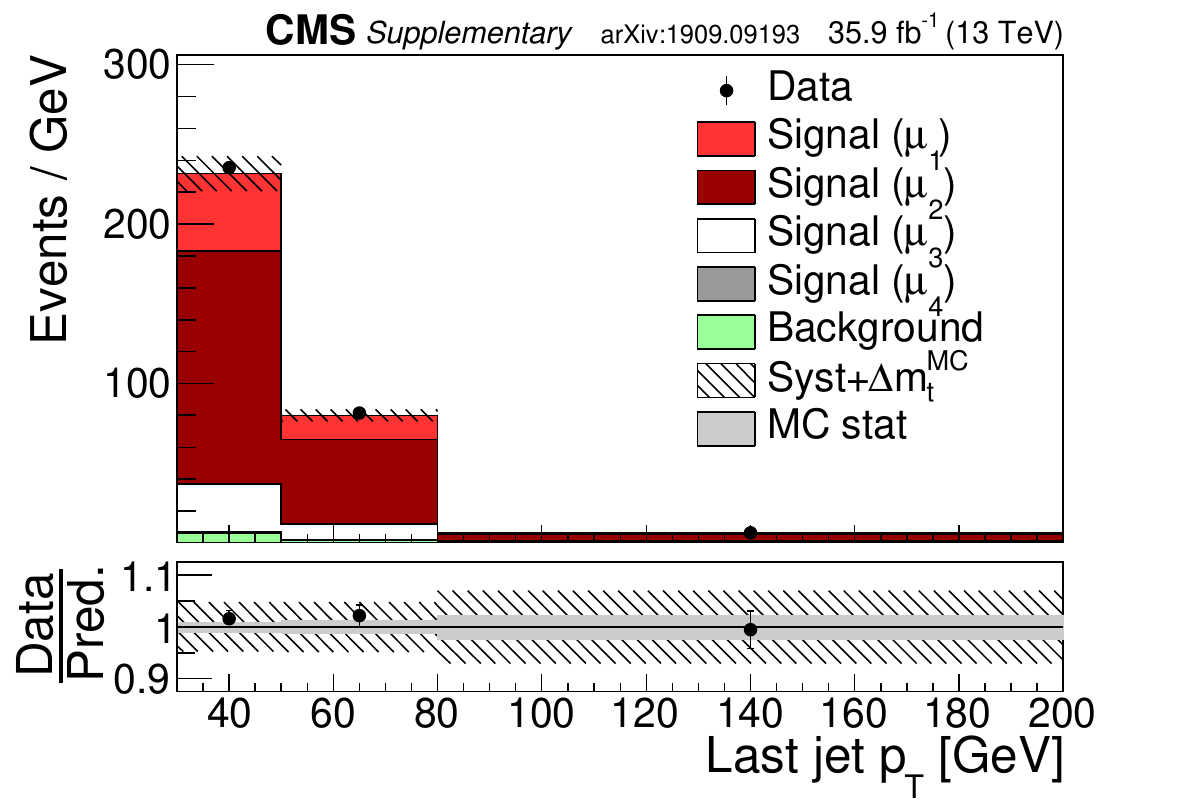}

    \includegraphics[width=0.325\textwidth]{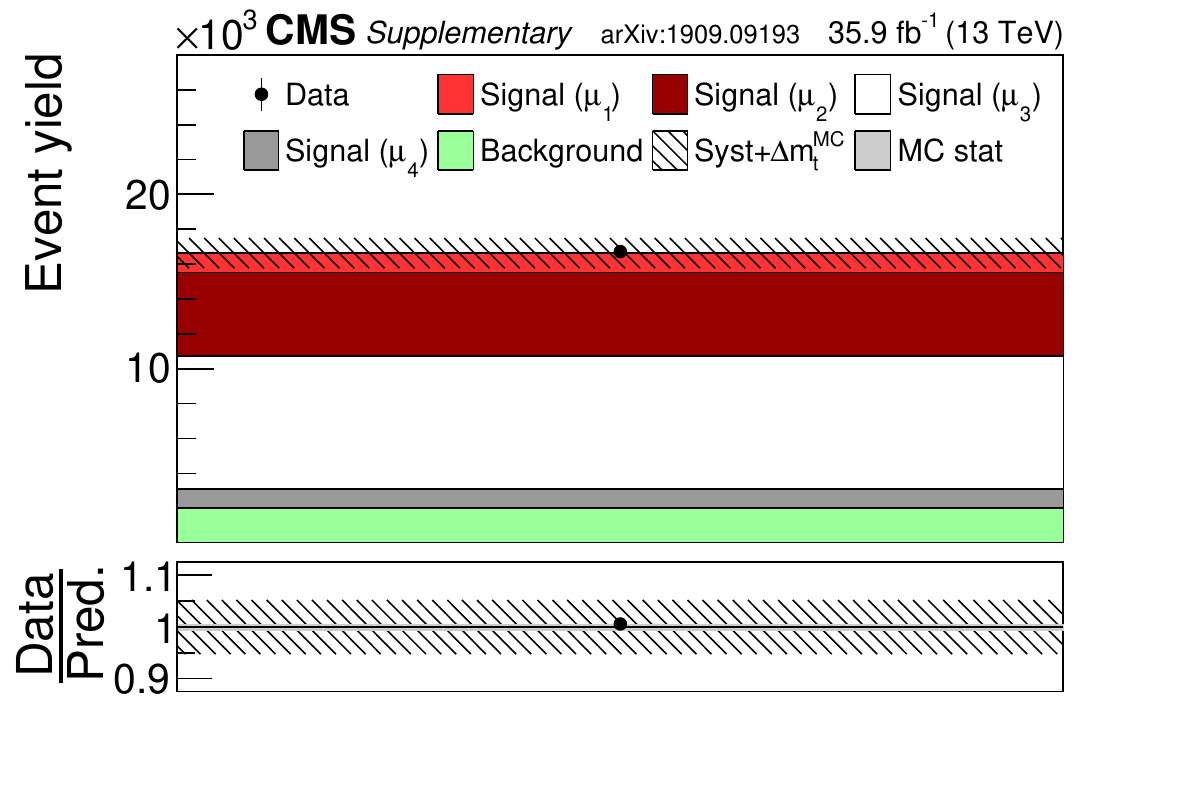}
    \includegraphics[width=0.325\textwidth]{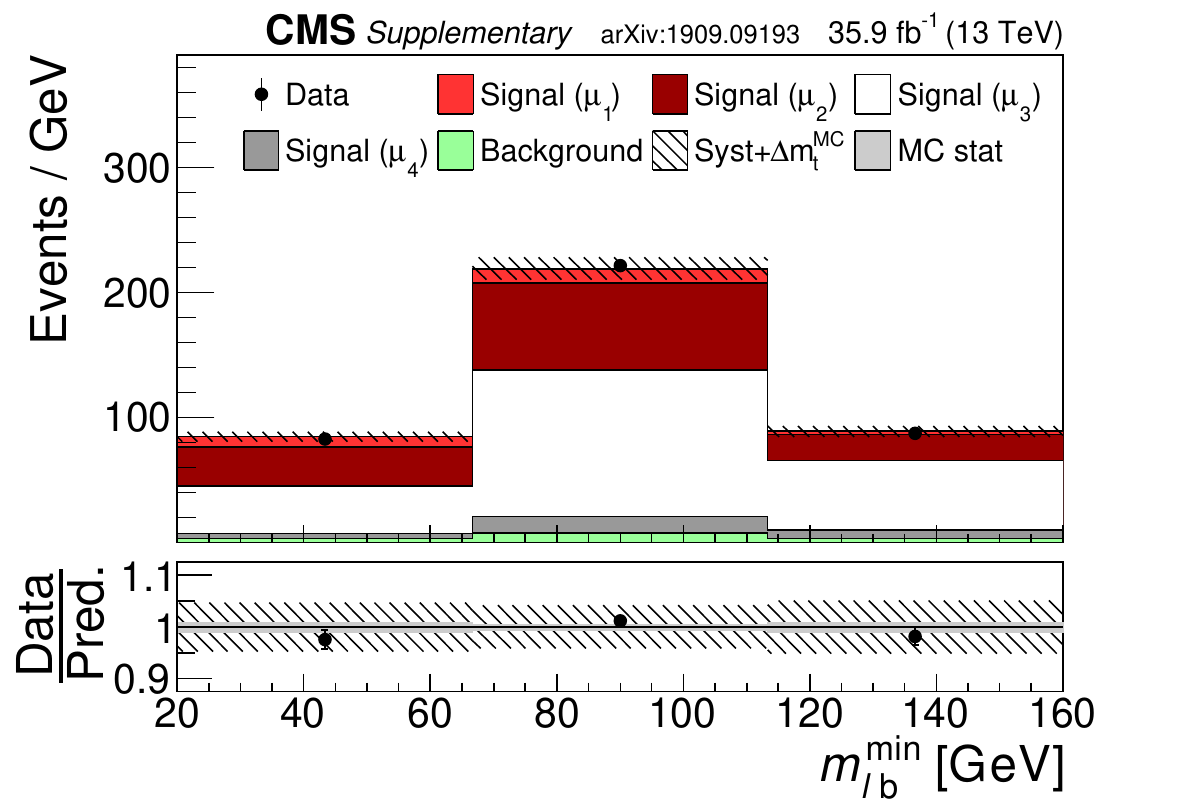}
    \includegraphics[width=0.325\textwidth]{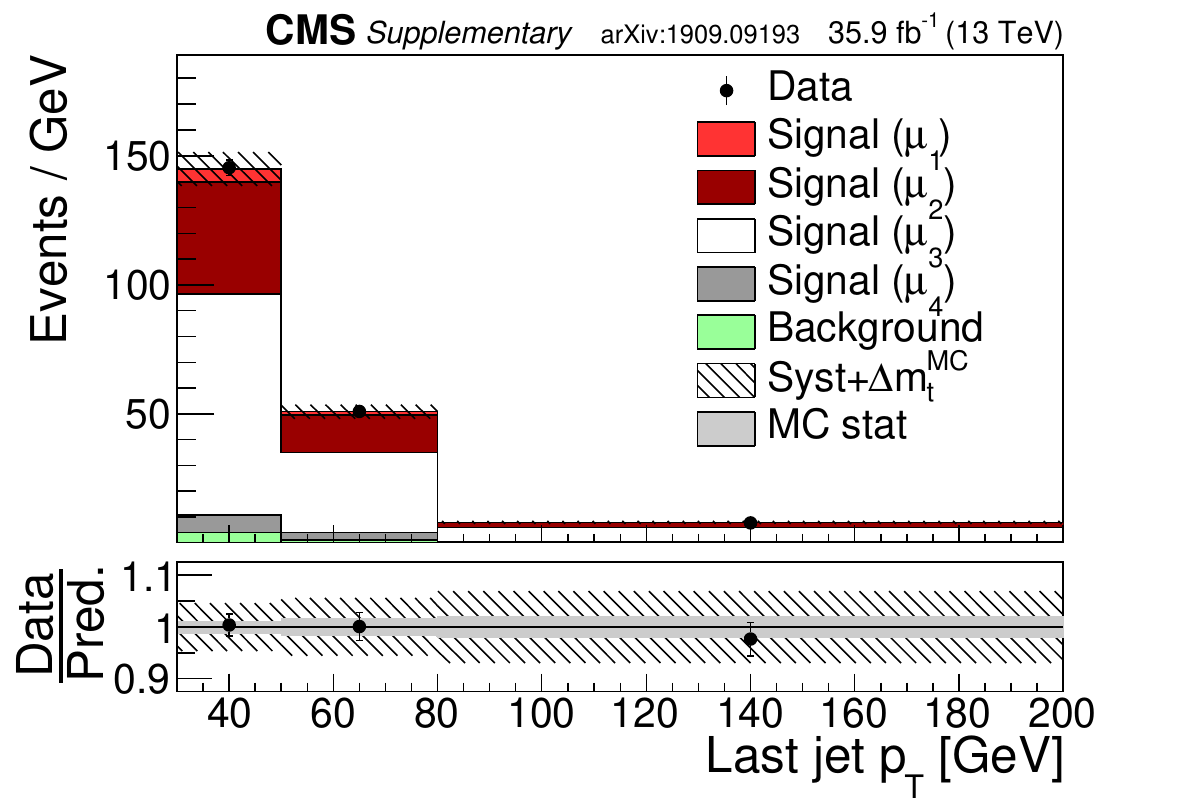}

    \includegraphics[width=0.325\textwidth]{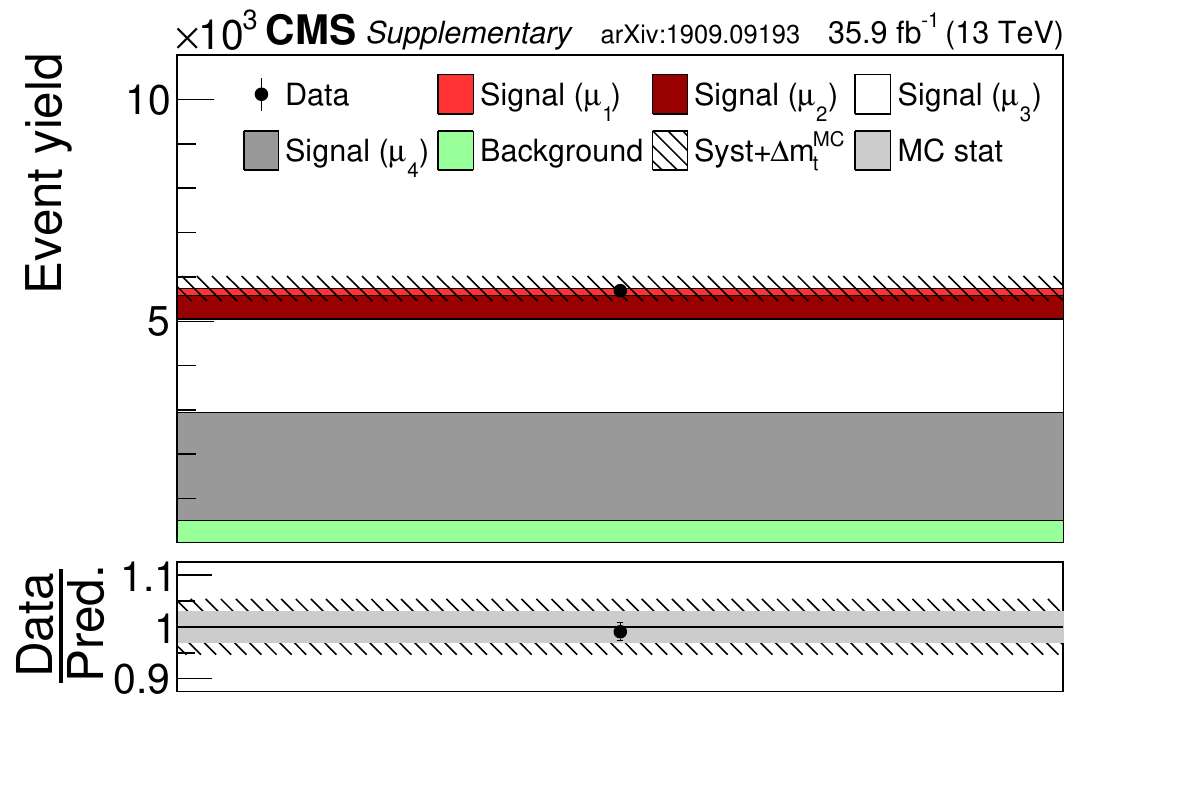}
    \includegraphics[width=0.325\textwidth]{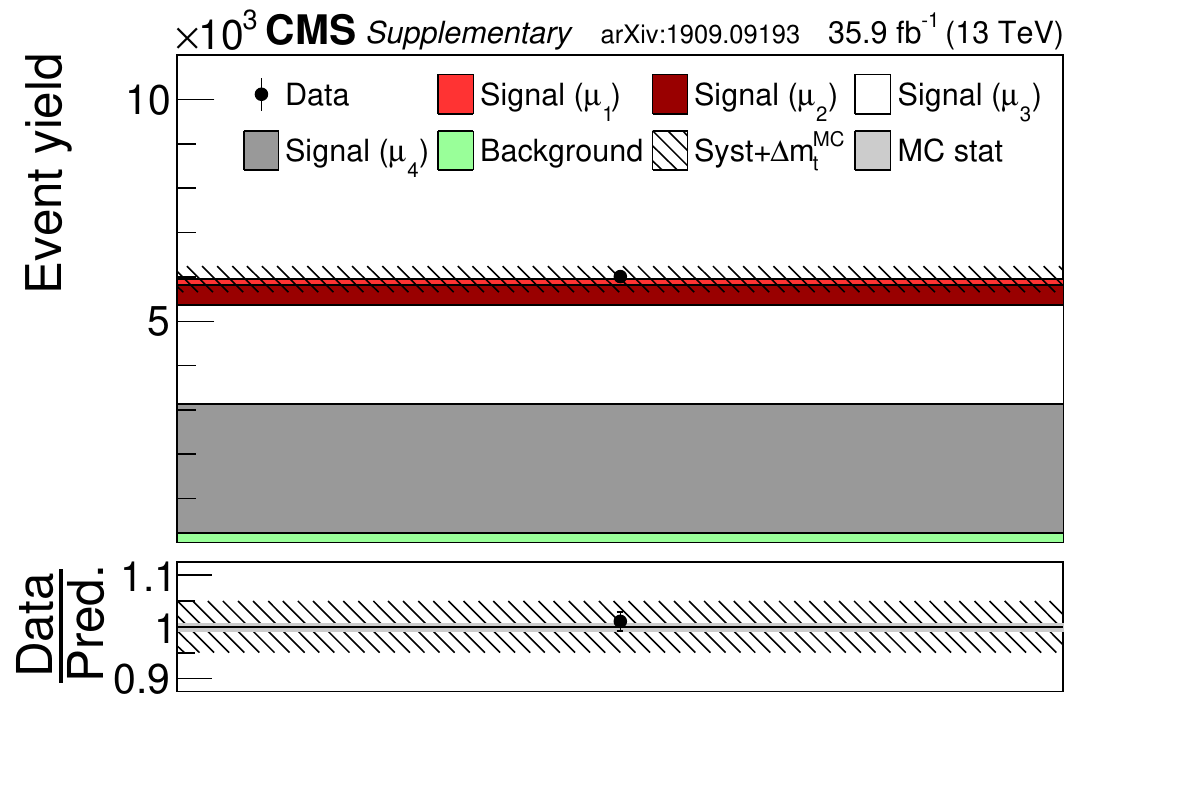}
    \includegraphics[width=0.325\textwidth]{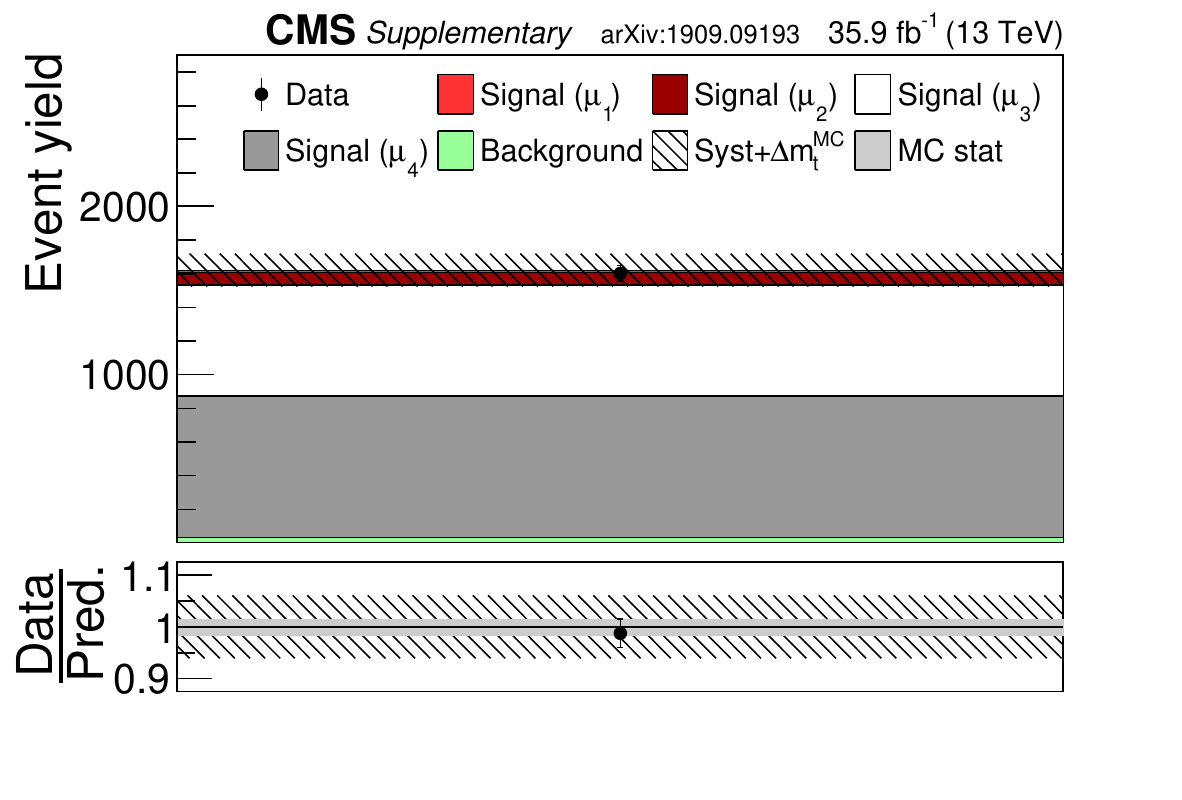}

    \includegraphics[width=0.325\textwidth]{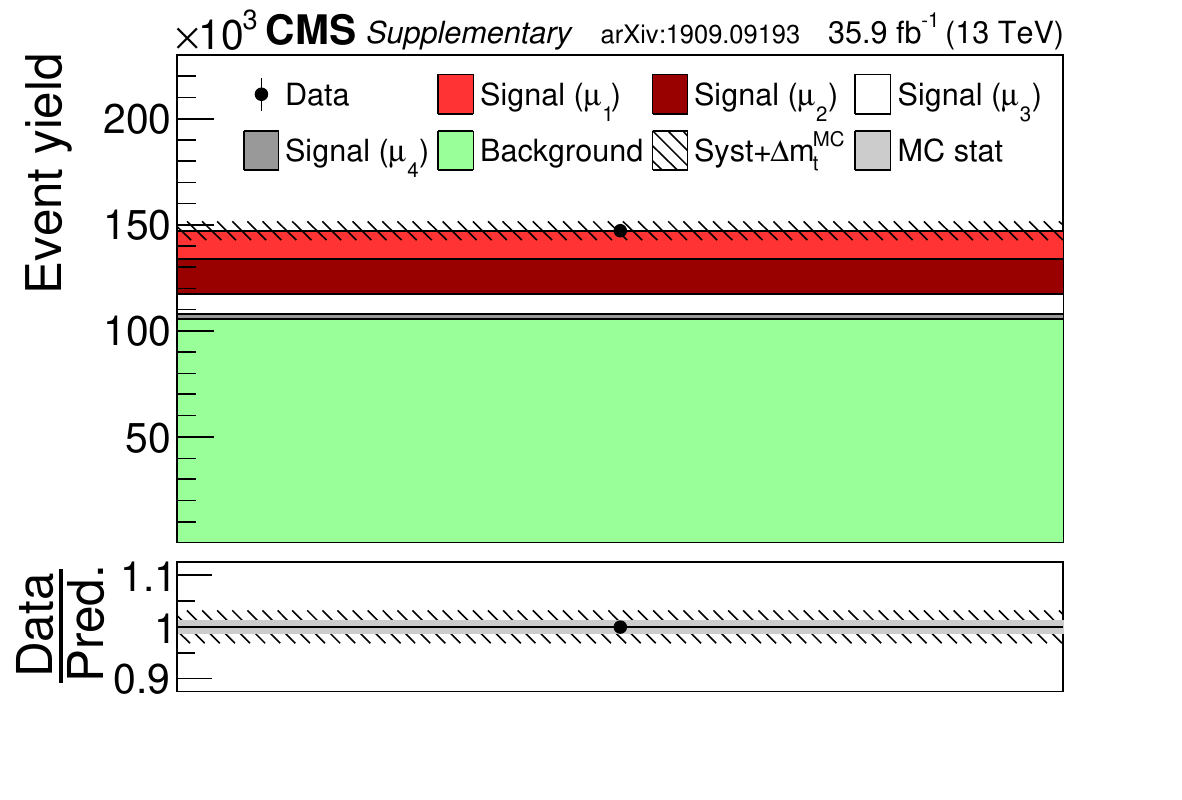}
    \includegraphics[width=0.325\textwidth]{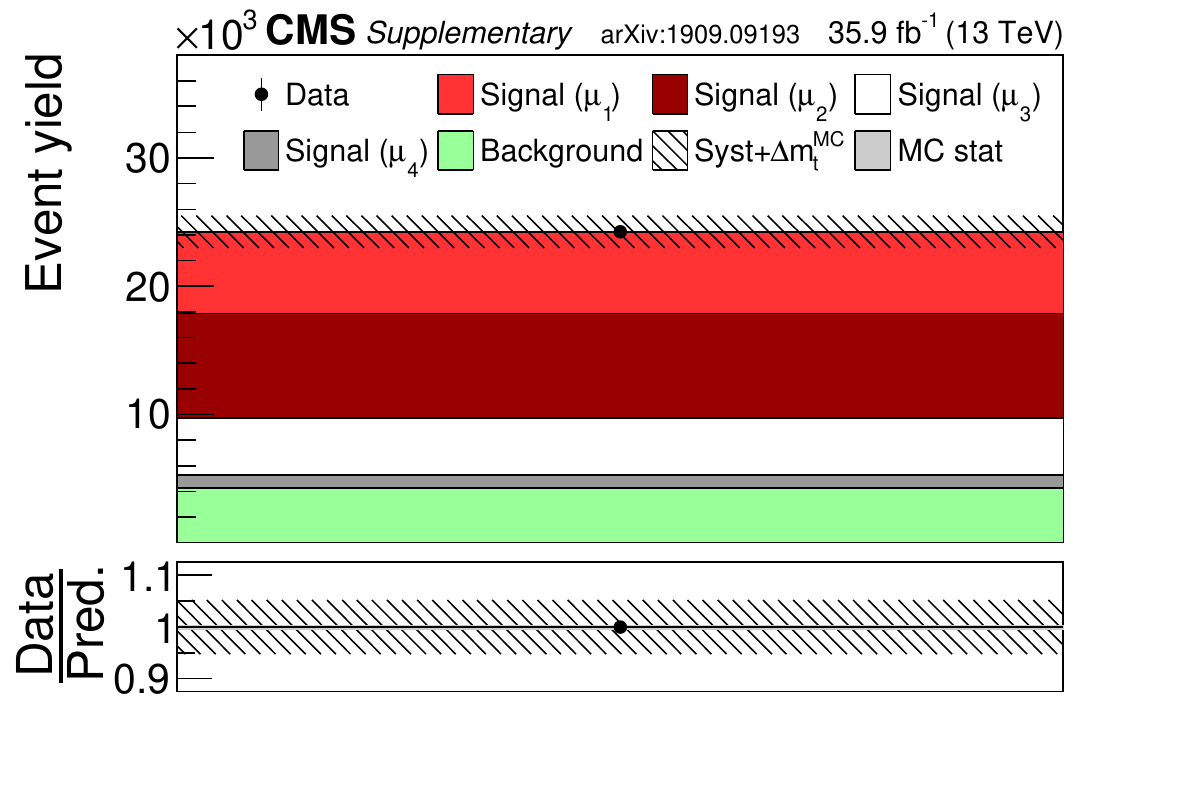}
    \hspace{0.325\textwidth}

    \caption{Comparison between data (points) and post-fit distributions of the expected signal and backgrounds from simulation
      (shaded histograms) used in the fit of \dstt. In the left column events with zero or three or more
      \cPqb-tagged jets are shown. The middle (right) column shows events with exactly one (two) \cPqb-tagged jets.
      Events in the first, second, third, and fourth bin of \mttreco are shown in the first, second, third, and fourth row, respectively,
      while events with less than two jets are shown in the fifth row.
      The hatched bands correspond to the total uncertainty in the sum of the predicted yields and include the contribution
      from the top quark mass ($\Delta\mtmc$). The ratios of data to the sum of the predicted yields are shown in the lower panel of each figure.
      Here, the solid gray band represents the contribution of the statistical uncertainty.
    \label{fig:postfit}}
\end{figure*}

\begin{figure*}[htbp]
  \centering
  \includegraphics[width=0.85\textwidth]{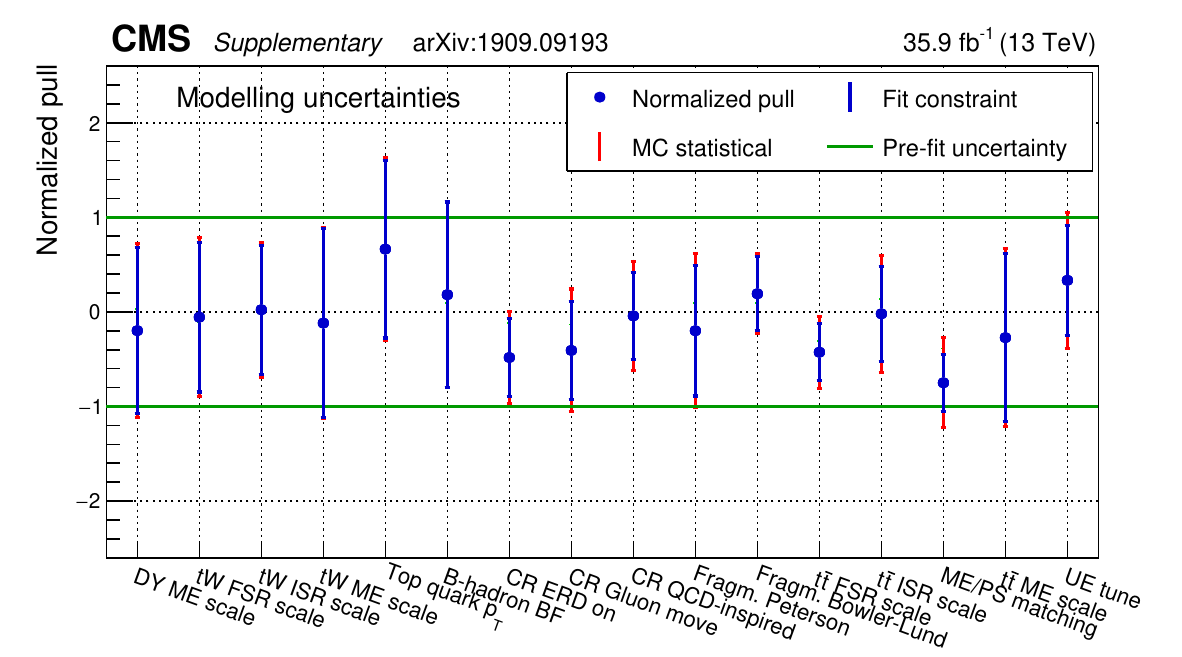}
  \caption{
    Pulls and constraints of the nuisance parameters related to the modelling uncertainties. The pulls, defined as the difference between the
    post-fit and pre-fit values of a nuisance parameter in units of the corresponding pre-fit uncertainty, are represented by the markers,
    while the constraints, defined as the ratio between the post-fit and pre-fit uncertainties on a nuisance parameter, correspond to the
    inner vertical bands. The horizontal lines represent the pre-fit uncertainty, and the outer vertical bands indicate
    the additional uncertainty due to the limited statistics in the simulation,
    as determined using pseudo-experiments.
    \label{fig:pulls}}
\end{figure*}

\begin{table}[tbh]
  \centering
  \topcaption{
    Correlations between the measured \sttk, including all systematic uncertainties.
    \label{tab:corr}}
  \begin{tabular}{l| cccc}
    & $\stt^{(\mu_1)}$ & $\stt^{(\mu_2)}$ & $\stt^{(\mu_3)}$ & $\stt^{(\mu_4)}$ \\ \hline
    $\stt^{(\mu_1)}$ & 1.00 &  & & \\ 
    $\stt^{(\mu_2)}$ & 0.64 & 1.00 & & \\ 
    $\stt^{(\mu_3)}$ & 0.72 & 0.60 & 1.00 & \\
    $\stt^{(\mu_4)}$ & 0.32 & 0.65 & 0.47 & 1.00 \\
    \end{tabular}
\end{table}

\begin{table}[tbh]
  \centering
  \topcaption{
   The relative uncertainty in $\stt^{(\mu_k)}$ and its sources, as obtained from the likelihood fit.
   The MC statistical uncertainty is determined separately, using pseudo-experiments.
   The individual uncertainties are given without their correlations, which are however accounted for in the total uncertainty.
   For extrapolation uncertainties, the $\pm$ notation is used if a positive variation produces
   an increase in $\stt^{(\mu_k)}$, while the $\mp$ notation is used otherwise.
      \label{tab:stt1}}
  \begin{tabular}{lc}
    Source & Uncertainty [\%] \\ \hline
    Jet energy scale & ${1.0}$ \\
    PDF & ${1.1}$ \\
    Lepton ID/isolation & ${2.2}$ \\
    Electron energy & ${0.5}$ \\
    \cPqb~quark fragmentation & ${1.1}$ \\
    \cPqb~tagging & ${0.2}$ \\
    Colour reconnection & ${0.7}$ \\
    Kinematic reconstruction & ${0.4}$ \\
    DY ME scale & ${0.4}$ \\
    Jet energy resolution & ${0.2}$ \\
    Muon energy scale & ${0.1}$ \\
    Pile-up & ${0.5}$ \\
    \tW FSR scale & ${0.2}$ \\
    \tW ISR scale & ${0.2}$ \\
    \tW ME scale & ${0.2}$ \\
    \mtmc & ${0.5}$ \\
    Top quark \pt & ${0.7}$ \\
    Trigger & ${0.3}$ \\
    \cPqb~hadron BF & ${0.1}$ \\
    \ttbar FSR scale & ${0.7}$ \\
    \ttbar ISR scale & ${0.3}$ \\
    ME/PS matching & ${0.2}$ \\
    \ttbar ME scale & ${0.3}$ \\
    UE tune & ${0.3}$ \\
    DY background & ${0.9}$ \\
    \tW background & ${0.6}$ \\
    \Wjets background & ${0.1}$ \\
    Diboson background & ${0.6}$ \\
    \ttbar background & ${0.3}$ \\
    Integrated luminosity & ${2.6}$ \\
    Statistical & ${0.7}$ \\
    MC statistical  & ${1.5}$ \\
    \multicolumn{2}{l}{Extrapolation uncertainties} \\ [0.05cm] \hline
    \rule{0pt}{2.3ex}\ttbar ISR scale & $\pm 0.2$ \\
    \rule{0pt}{2.3ex}\ttbar FSR scale & $\pm 0.1$ \\
    \rule{0pt}{2.3ex}\ttbar ME scale & $\pm 0.1$ \\
    \rule{0pt}{2.3ex}UE tune & $\mp^{<0.1}_{0.1}$ \\
    \rule{0pt}{2.3ex}PDF & $\pm^{0.8}_{0.5}$ \\
    \rule{0pt}{2.3ex}Top quark \pt & $\pm^{<0.1}_{0.1}$ \\ 
    \rule{0pt}{2.5ex}Total $\stt^{(\mu_1)}$ uncertainty & $^{+4.7}_{-4.4}$ \\ 
  \end{tabular}
\end{table}

\begin{table}[tbh]
  \centering
  \topcaption{
    Same as Table~\ref{tab:stt1}, but for $\stt^{(\mu_2)}$.
    \label{tab:stt2}}
  \begin{tabular}{lc}
    Source & Uncertainty [\%] \\ \hline
    Jet energy scale & ${1.3}$ \\
    PDF & ${1.0}$ \\
    Lepton ID/isolation & ${2.3}$ \\
    Electron energy & ${0.4}$ \\
    \cPqb~quark fragmentation & ${1.0}$ \\
    \cPqb~tagging & ${0.5}$ \\
    Colour reconnection & ${1.0}$ \\
    Kinematic reconstruction & ${0.4}$ \\
    DY ME scale & ${0.2}$ \\
    Jet energy resolution & ${0.5}$ \\
    Muon energy scale & ${0.2}$ \\
    Pile-up & ${0.2}$ \\
    \tW FSR scale & ${0.2}$ \\
    \tW ISR scale & ${0.2}$ \\
    \tW ME scale & ${0.2}$ \\
    \mtmc & ${0.4}$ \\
    Top quark \pt & ${0.4}$ \\
    Trigger & ${0.4}$ \\
    \cPqb~hadron BF & ${0.2}$ \\
    \ttbar FSR scale & ${1.5}$ \\
    \ttbar ISR scale & ${0.3}$ \\
    ME/PS matching & ${0.8}$ \\
    \ttbar ME scale & ${0.8}$ \\
    UE tune & ${0.2}$ \\
    DY background & ${1.2}$ \\
    \tW background & ${1.1}$ \\
    \Wjets background & ${0.2}$ \\
    Diboson background & ${0.3}$ \\
    \ttbar background & ${0.2}$ \\
    Integrated luminosity & ${2.6}$ \\
    Statistical & ${0.6}$ \\
    MC statistical  & ${1.8}$ \\
    \multicolumn{2}{l}{Extrapolation uncertainties} \\ [0.05cm] \hline
    \rule{0pt}{2.3ex}\ttbar ISR scale & $\mp^{0.2}_{0.1}$ \\
    \rule{0pt}{2.3ex}\ttbar FSR scale & $\mp^{<0.1}_{0.1}$ \\
    \rule{0pt}{2.3ex}\ttbar ME scale & $\mp^{0.1}_{0.2}$ \\
    \rule{0pt}{2.3ex}UE tune & $\mp^{0.1}_{<0.1}$ \\
    \rule{0pt}{2.3ex}PDF & $\pm^{0.8}_{0.6}$ \\
    \rule{0pt}{2.3ex}Top quark \pt & $\pm 0.1$ \\ 
    \rule{0pt}{2.5ex}Total $\stt^{(\mu_2)}$ uncertainty & $^{+5.0}_{-4.8}$ \\ 
  \end{tabular}
\end{table}

\begin{table}[tbh]
  \centering
  \topcaption{
    Same as Table~\ref{tab:stt1}, but for $\stt^{(\mu_3)}$.
    \label{tab:stt3}}
  \begin{tabular}{lc}
    Source & Uncertainty [\%] \\ \hline
    Jet energy scale & ${1.8}$ \\
    PDF & ${1.2}$ \\
    Lepton ID/isolation & ${2.2}$ \\
    Electron energy & ${0.5}$ \\
    \cPqb~quark fragmentation & ${1.1}$ \\
    \cPqb~tagging & ${0.7}$ \\
    Colour reconnection & ${0.7}$ \\
    Kinematic reconstruction & ${0.3}$ \\
    DY ME scale & ${0.1}$ \\
    Jet energy resolution & ${0.1}$ \\
    Muon energy scale & ${0.1}$ \\
    Pile-up & ${0.3}$ \\
    \tW FSR scale & ${0.1}$ \\
    \tW ISR scale & ${0.1}$ \\
    \tW ME scale & ${0.1}$ \\
    \mtmc & ${0.6}$ \\
    Top quark \pt & ${1.2}$ \\
    Trigger & ${0.4}$ \\
    \cPqb~hadron BF & ${0.1}$ \\
    \ttbar FSR scale & ${0.5}$ \\
    \ttbar ISR scale & ${0.5}$ \\
    ME/PS matching & ${0.5}$ \\
    \ttbar ME scale & ${0.7}$ \\
    UE tune & ${0.2}$ \\
    DY background & ${1.2}$ \\
    \tW background & ${0.9}$ \\
    \Wjets background & ${0.1}$ \\
    Diboson background & ${0.3}$ \\
    \ttbar background & ${0.1}$ \\
    Integrated luminosity & ${2.6}$ \\
    Statistical & ${0.8}$ \\
    MC statistical  & ${1.4}$ \\
    \multicolumn{2}{l}{Extrapolation uncertainties} \\ [0.05cm] \hline
    \rule{0pt}{2.3ex}\ttbar ISR scale & $\pm^{0.2}_{0.1}$ \\
    \rule{0pt}{2.3ex}\ttbar FSR scale & $\pm^{0.2}_{<0.1}$ \\
    \rule{0pt}{2.3ex}\ttbar ME scale & $\mp^{0.4}_{0.5}$ \\
    \rule{0pt}{2.3ex}UE tune & $\pm 0.1$ \\
    \rule{0pt}{2.3ex}PDF & $\pm^{0.9}_{0.6}$ \\
    \rule{0pt}{2.3ex}Top quark \pt & $\pm^{0.2}_{0.4}$ \\ 
    \rule{0pt}{2.5ex}Total $\stt^{(\mu_3)}$ uncertainty & $^{+5.0}_{-4.8}$ \\ 
  \end{tabular}
\end{table}

\begin{table}[tbh]
  \centering
  \topcaption{
    Same as Table~\ref{tab:stt1}, but for $\stt^{(\mu_4)}$.
    \label{tab:stt4}}
  \begin{tabular}{lc}
    Source & Uncertainty [\%] \\ \hline
    Jet energy scale & ${2.2}$ \\
    PDF & ${1.5}$ \\
    Lepton ID/isolation & ${2.0}$ \\
    Electron energy & ${0.4}$ \\
    \cPqb~quark fragmentation & ${0.9}$ \\
    \cPqb~tagging & ${1.2}$ \\
    Colour reconnection & ${1.4}$ \\
    Kinematic reconstruction & ${0.6}$ \\
    DY ME scale & ${0.4}$ \\
    Jet energy resolution & ${0.8}$ \\
    Muon energy scale & ${0.4}$ \\
    Pile-up & ${0.4}$ \\
    \tW FSR scale & ${0.4}$ \\
    \tW ISR scale & ${0.4}$ \\
    \tW ME scale & ${0.4}$ \\
    \mtmc & ${0.4}$ \\
    Top quark \pt & ${3.5}$ \\
    Trigger & ${0.5}$ \\
    \cPqb~hadron BF & ${0.4}$ \\
    \ttbar FSR scale & ${1.3}$ \\
    \ttbar ISR scale & ${0.4}$ \\
    ME/PS matching & ${1.0}$ \\
    \ttbar ME scale & ${1.8}$ \\
    UE tune & ${0.8}$ \\
    DY background & ${1.6}$ \\
    \tW background & ${1.0}$ \\
    \Wjets background & ${0.5}$ \\
    Diboson background & ${0.4}$ \\
    \ttbar background & ${0.4}$ \\
    Integrated luminosity & ${2.6}$ \\
    Statistical & ${1.8}$ \\
    MC statistical  & ${2.5}$ \\
    \multicolumn{2}{l}{Extrapolation uncertainties} \\ [0.05cm] \hline
    \rule{0pt}{2.3ex}\ttbar ISR scale & $\mp^{0.8}_{0.7}$ \\
    \rule{0pt}{2.3ex}\ttbar FSR scale & $\pm^{0.2}_{<0.1}$ \\
    \rule{0pt}{2.3ex}\ttbar ME scale & $\mp^{0.8}_{1.2}$ \\
    \rule{0pt}{2.3ex}UE tune & $\pm^{0.1}_{0.2}$ \\
    \rule{0pt}{2.3ex}PDF & $\pm^{1.2}_{0.9}$ \\
    \rule{0pt}{2.3ex}Top quark \pt & $\pm^{0.6}_{1.2}$ \\ 
    \rule{0pt}{2.5ex}Total $\stt^{(\mu_4)}$ uncertainty & $^{+7.2}_{-6.9}$ \\ 
  \end{tabular}
\end{table}

\begin{figure}[htbp]
  \centering
  \includegraphics[width=0.5\textwidth]{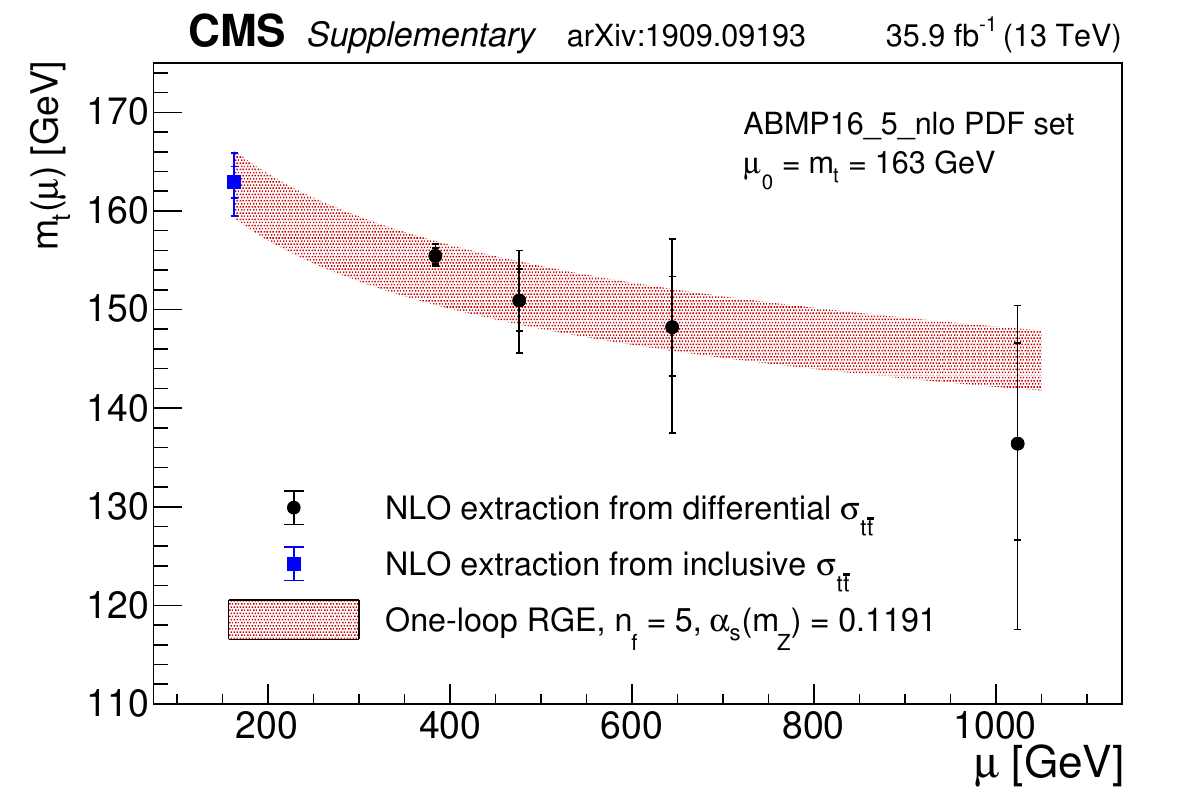}
  \caption{Values of $\mt(\mu_k)$ extracted from the measured \dstt (round markers), compared to the value of $\mt(\mt)$ extracted
    from the inclusive \stt (square marker). The inner vertical bands correspond to the combination of fit, extrapolation, and PDF uncertainties,
    while the outer vertical bands include the contribution of the scale uncertainties.
  \label{fig:mtmuk}}
\end{figure}

  }
\cleardoublepage \section{The CMS Collaboration \label{app:collab}}\begin{sloppypar}\hyphenpenalty=5000\widowpenalty=500\clubpenalty=5000\vskip\cmsinstskip
\textbf{Yerevan Physics Institute, Yerevan, Armenia}\\*[0pt]
A.M.~Sirunyan$^{\textrm{\dag}}$, A.~Tumasyan
\vskip\cmsinstskip
\textbf{Institut f\"{u}r Hochenergiephysik, Wien, Austria}\\*[0pt]
W.~Adam, F.~Ambrogi, T.~Bergauer, J.~Brandstetter, M.~Dragicevic, J.~Er\"{o}, A.~Escalante~Del~Valle, M.~Flechl, R.~Fr\"{u}hwirth\cmsAuthorMark{1}, M.~Jeitler\cmsAuthorMark{1}, N.~Krammer, I.~Kr\"{a}tschmer, D.~Liko, T.~Madlener, I.~Mikulec, N.~Rad, J.~Schieck\cmsAuthorMark{1}, R.~Sch\"{o}fbeck, M.~Spanring, D.~Spitzbart, W.~Waltenberger, C.-E.~Wulz\cmsAuthorMark{1}, M.~Zarucki
% ...
\vskip\cmsinstskip
\dag: Deceased\\
1:  Also at Vienna University of Technology, Vienna, Austria\\
% ...
\end{sloppypar}
\end{document}